\documentclass[11pt,a4paper]{article}
\usepackage[a4paper,left=1in,right=1in,top=1.3in,bottom=1.3in]{geometry}
\usepackage{graphicx}
\usepackage{amsmath}
\usepackage{amssymb}
\usepackage{xspace}
\usepackage{cite}
\usepackage{url}
\usepackage{hyperref}

\newcommand{\nbar}{\ensuremath{\bar{n}}}
\newcommand{\LO}{\text{LO}\xspace}
\newcommand{\nLO}{\nbar \text{LO}\xspace}
\newcommand{\nnLO}{\nbar \nbar \text{LO}\xspace}
\newcommand{\nNLO}{\nbar \text{NLO}\xspace}
\newcommand{\nnNLO}{\nbar \nbar \text{NLO}\xspace}
\newcommand{\NNLO}{\text{NNLO}\xspace}
\newcommand{\NLO}{\text{NLO}\xspace}
\newcommand{\GeV}{\ensuremath{\,\mathrm{GeV}}\xspace}
\newcommand{\TeV}{\ensuremath{\,\mathrm{TeV}}\xspace}
\newcommand{\fb}{\ensuremath{\,\mathrm{fb}}\xspace}
\newcommand{\pb}{\ensuremath{\,\mathrm{pb}}\xspace}
\newcommand{\jet}{\ensuremath{\,\mathrm{jet}}\xspace}
\newcommand{\tot}{\ensuremath{\,\mathrm{tot}}\xspace}
\newcommand{\LS}{{\ensuremath{\,\mathrm{LS}}\xspace}}
\newcommand{\miss}{\ensuremath{\,\mathrm{miss}}\xspace}
\newcommand{\order}[1]{\mathcal{O}\!\left(#1\right)}
\newcommand{\as}{\alpha_s}
\newcommand{\aEW}{\alpha_\text{\sc ew}}
\newcommand{\mcfm}{\textsc{mcfm}}
\newcommand{\bh}{\textsc{blackhat}}
\newcommand{\sherpa}{\textsc{sherpa}}
\newcommand{\alpgen}{\textsc{alpgen}}
\newcommand{\pythia}{\textsc{pythia}}
\newcommand{\herwig}{\textsc{herwig}}
\newcommand{\roo}{\textsc{root}}
\newcommand{\rocket}{\textsc{rocket}}
\newcommand{\ie}{i.e.}

\title{Simulated NNLO for high-$p_T$ observables  \\
in vector boson + jets production at the LHC}

\author{
  Daniel Ma\^{i}tre and Sebastian Sapeta \\ \\
  {\it Institute for Particle Physics Phenomenology, Durham University,}\\
  {\it South Rd, Durham DH1 3LE, UK}
}
\date{}

\begin{document}
\maketitle
\vspace{-20em}
\begin{flushright}
  IPPP/13/49\\
  DCPT/13/98
\end{flushright}
\vspace{15em}

\begin{abstract}
We present a study of higher order QCD corrections beyond NLO to processes with
an electroweak vector boson, W or Z, in association with jets. We focus on the
regions of high transverse momenta of commonly used differential distributions.
We employ the LoopSim method to merge NLO samples of different multiplicity
obtained from \mcfm~and from \bh+\sherpa~in order to compute the dominant
part of the NNLO corrections for high-$p_T$ observables.
We find that these corrections are indeed substantial for a number of
experimentally relevant observables. For other observables, they lead to
significant reduction of scale uncertainties.
\end{abstract}

\section{Introduction}

The production of electroweak vector boson in association with jets forms one of
the most studied class of Standard Model (SM) processes.
W + jets is a background to single and pair top production, and both W and Z
in association with jets constitute significant backgrounds to processes with
dibosons, Higgs production as well as to searches for physics beyond the
standard model (BSM). 
V+jets processes are also interesting in their own right, as they
provide important tests of quantum chromodynamics (QCD).
 
Both W+jets and Z+jets production has been recently studied experimentally at
the LHC, $\sqrt{s} = 7 \TeV$, by the ATLAS~\cite{Aad:2011fp,
Aad:2011qv,Aad:2012en, Aad:2013ysa} and CMS~\cite{Chatrchyan:2011ne,
Chatrchyan:2013tna} collaborations. 

There exist a range of theoretical frameworks and tools that allow one to make
predictions of cross sections and distributions for the W/Z+jets processes.
Many of them have been used in the context of the LHC.
Results at fixed next-to-leading-order (NLO) in QCD for W/Z in association with
up to 2 jets are available from \mcfm~\cite{mcfm, Campbell:2003hd}, up to 3 jets
from \rocket~\cite{EMZW3j}, and up to 4 jets (5 in the case of W) from
\bh~\cite{BLACKHAT}+\sherpa~\cite{SHERPA}.  The latter has been used as well to
study NLO samples merged with the Exclusive Sums
method~\cite{AlcarazMaestre:2012vp}.
V+jets production at the LHC has been also studied with the traditional LO Monte
Carlo (MC) programs including the parton shower (PS) and hadronization stages,
\pythia~\cite{Sjostrand:2006za}, \herwig~\cite{Corcella:2000bw},
\sherpa~\cite{SHERPA}, \alpgen~\cite{Mangano:2002ea}, the latter allowing for
merging of LO samples with different multiplicities.
The HEJ formalism~\cite{HEJ}, based on high-energy resummation, has
been also used to study the production of a W boson associated with
jets~\cite{Andersen:2012gk}.
Finally, recent years have seen an enormous progress in the field of NLO + PS
matching and NLO merging, with a range of new techniques being used to study
V+jets production. Those started with MC@NLO~\cite{Frixione:2002ik} and
POWHEG~\cite{Alioli:2010xd} and were further developed and refined into new
methods: MEPS@NLO~\cite{Hoeche:2012ft,Hoeche:2012yf},
MiNLO~\cite{Hamilton:2012np, Hamilton:2012rf}, FxFx~\cite{Frederix:2012ps}, 
UNLOPS~\cite{Lonnblad:2012ix} and a similar unitarity-preserving
approach by Platzer~\cite{Platzer:2012bs}. The two last methods are both related
to LoopSim~\cite{Rubin:2010xp}.

In this paper, we present a study of W/Z+jets processes at approximate
next-to-next-to-leading order (NNLO) in QCD, in the context of the LHC,
$\sqrt{s} = 7 \TeV$.
Our NNLO results include exact double-real and real-virtual
contributions as well as exact singular terms of the 2-loop diagrams.
The motivation to go beyond NLO comes from the fact that the next-to-leading
order corrections for these processes turn out to be sizable for a number of
important distributions at high transverse momentum.
These corrections come about due to new production channels and new topologies
absent at LO and appearing for the first time at NLO. 
For example, at leading order, the production of a vector boson in
association with a jet is possible only via $q\bar q$ or $qg$ channels. At NLO,
the new $qq$ channel, with enhanced partonic luminosity, opens up
and adds a substantial contribution to the cross section. 
Similarly, at LO, only back-to-back V+jet configurations are possible. 
Henceforth, we shall call them the ``LO-type topologies''.
At NLO, however, a totally different type of topology appears, with two hard
QCD partons recoiling against each other and the electroweak boson, emitted from
a quark line, being soft or collinear. The latter brings logarithmic
enhancements for a number of distributions. In the following, we shall call such
configurations ``dijet-type topologies''.

Because the NLO corrections often turn out to be commensurate with the leading
order, it is of great importance to try to assess the NNLO contributions, to
check the convergence of the perturbative series, and to obtain precise and
stable results. 
While the exact $\order{\aEW\as^3}$ results for the inclusive W/Z+jet production
are still missing, we can compute the dominant part of the NNLO corrections to
these processes at high $p_T$, using the LoopSim procedure to merge NLO samples with different multiplicities.

The LoopSim method was proposed in~\cite{Rubin:2010xp} and validated there in
the studies of Z+jets, Drell-Yan and dijets.
It has been also shown to successfully describe the Tevatron
data~\cite{Camarda:2013oej} and used for predictions of the WZ
production~\cite{Campanario:2012fk}.
Here we use it in the context of the experimental studies of Z+jets and W+jets
production at the LHC, $\sqrt{s} = 7 \TeV$~\cite{Aad:2011fp,
Aad:2011qv,Aad:2012en, Aad:2013ysa,Chatrchyan:2011ne}, following the ATLAS cuts,
and confronting our results with available data.
The method is briefly summarized in the following section. In order to
distinguish our predictions with simulated loops from those with exact loop
diagrams, we denote the approximate loops by $\nbar$, as opposed to N used for
the exact ones. So, for example, $\nLO$ means the correction with simulated
1-loop diagrams, but $\nNLO$ is a result with exact 1-loop and simulated 2-loop
contributions. Similarly, $\nnLO$ corresponds to the result with simulated
1-loop and simulated 2-loop diagrams and $\nnNLO$ would have exact 1-loop and
simulated 2 and 3-loop contributions.

The paper is organized as follows. In the next section, we provide a short
description of the LoopSim method and give details of our calculation. In
section ~\ref{sec:Vjetsresults}, we present the results for W+jets (Sec.
\ref{sec:wjets}) and Z+jets (Sec.  \ref{sec:zjets}) at \nNLO. We show
distributions for a range of important observables and, where possible, we
confront our predictions with the experimental data. In section
\ref{sec:ratios}, we discuss the ratios of differential distributions for either
W+jets and Z+jets or W$^+$+jets and W$^-$+jets, and comment on their
potential advantages. Finally, we summarise our study in
section~\ref{sec:conclusion}.

\section{Details of the calculation}
\label{sec:details}

For the calculation of W/Z+jets at \nNLO, we used LoopSim together with
\mcfm~\cite{mcfm, Campbell:2003hd} and independently with
\bh+\sherpa~ntuples.
The LoopSim method~\cite{Rubin:2010xp} allows for merging of the NLO samples
with different multiplicities to obtain approximate NNLO results. 
In the case of the V+jets process, the computation proceeds as follows: The NLO
program provides V+1j and V+2j weighted events. LoopSim takes these events and
starts by assigning a diagram to each of them. 
This is done by clustering the event with the Cambridge/Aachen~(C/A) jet
algorithm~\cite{Dokshitzer:1997in,Wobisch:1998wt}, as implemented in
FastJet~\cite{Cacciari:2005hq, FastJet}, with a certain radius $R_\LS$.
The C/A algorithm sequentially clusters pairs of particles closest in angle, that
is with the smallest $d_{ij} = \Delta R_{ij}^2/R_\LS^2$ measure, where
$\Delta R_{ij}^2 = (y_i-y_j)^2 + (\phi_i - \phi_j )^2$ is a distance between
particles $i$ and $j$ in the
rapidity-azimuthal angle plane. The indices $i,j = 1\dots n$, with $n$ being
the number of particles in the event. The algorithm stops when $d_{ij} \leq 1$
and all the remaining particles are clustered with the beam.
Then, each merging of partons $ij \to k$, done by the C/A algorithm, is
reinterpreted as splitting $k \to ij$ in the corresponding Feynman diagram. 
This interpretation is valid in the soft and collinear limit, where LoopSim
serves its main function of cancelling divergences.
 
In the next step, the underlying hard structure of the event is determined by
working through the $ij\to k$ recombinations in order of decreasing hardness,
defined by the $k_t$ algorithm~\cite{Kt, Ellis:1993tq} measure: $h_{ij} =
\min(p_{ti}^2,p_{tj}^2)\Delta R^2_{ij}/ R_\LS^2$ for $ij\to k$ merging and
$h_{iB} = p_{ti}^2$ for beam recombination. 
The first $b$ particles associated with the hardest merging are marked as
``Born''. In the case of V+jets, there are always two of them, either a boson
and a parton or two partons. The remaining non-Born particles are then
``looped'' by finding all possible ways of recombining them with the emitters.
In this step, LoopSim generates an approximate set of 1 and 2-loop diagrams with
the weights equal to $(-1)^\text{number of loops} \times \text{weight of the
original event}$. In the last step, a double counting, between the approximate
1-loop events generated by LoopSim and the exact 1-loop events coming from the
NLO sample with lower multiplicity, is removed. 
This is done essentially by generating 1-loop diagrams from the tree level
events first, and then using them to generate all possible 1 and 2-loop events.
This set is subtracted from the result of applying LoopSim to pure tree
level diagrams, which has 0, 1 and 2 approximate loops. For more details,
see~\cite{Rubin:2010xp}.

\begin{figure}[t]
  \includegraphics[width=0.48\columnwidth]{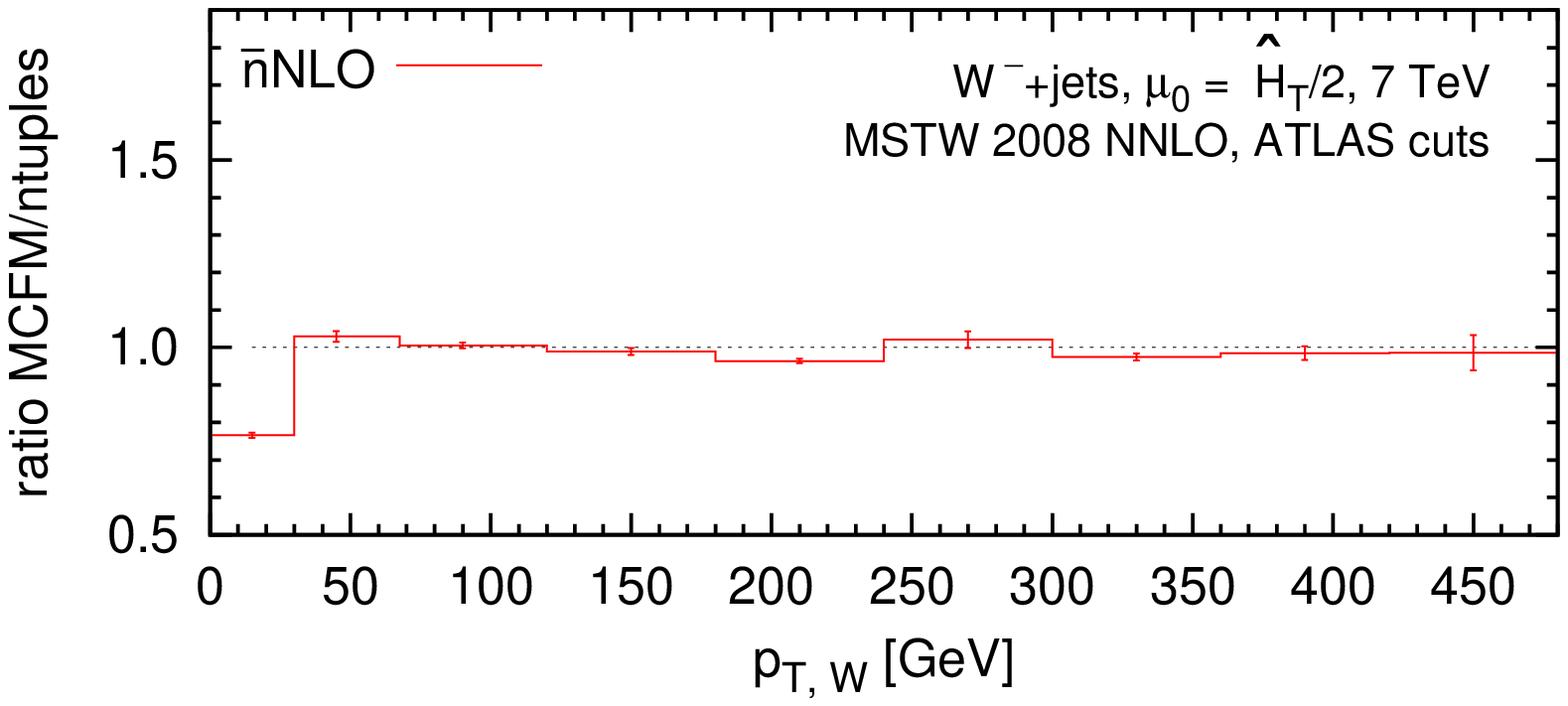}
  \hfill
  \includegraphics[width=0.48\columnwidth]{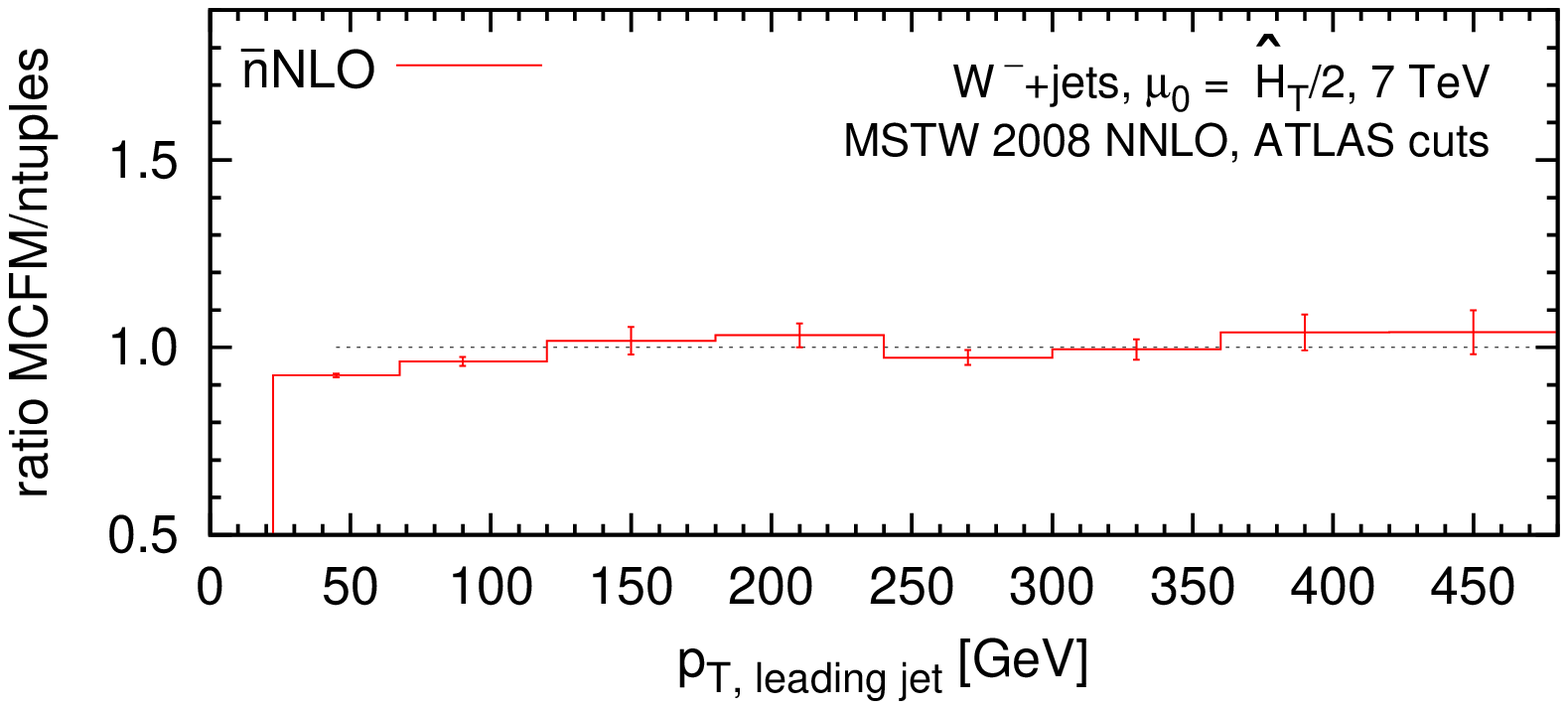}
  \includegraphics[width=0.48\columnwidth]{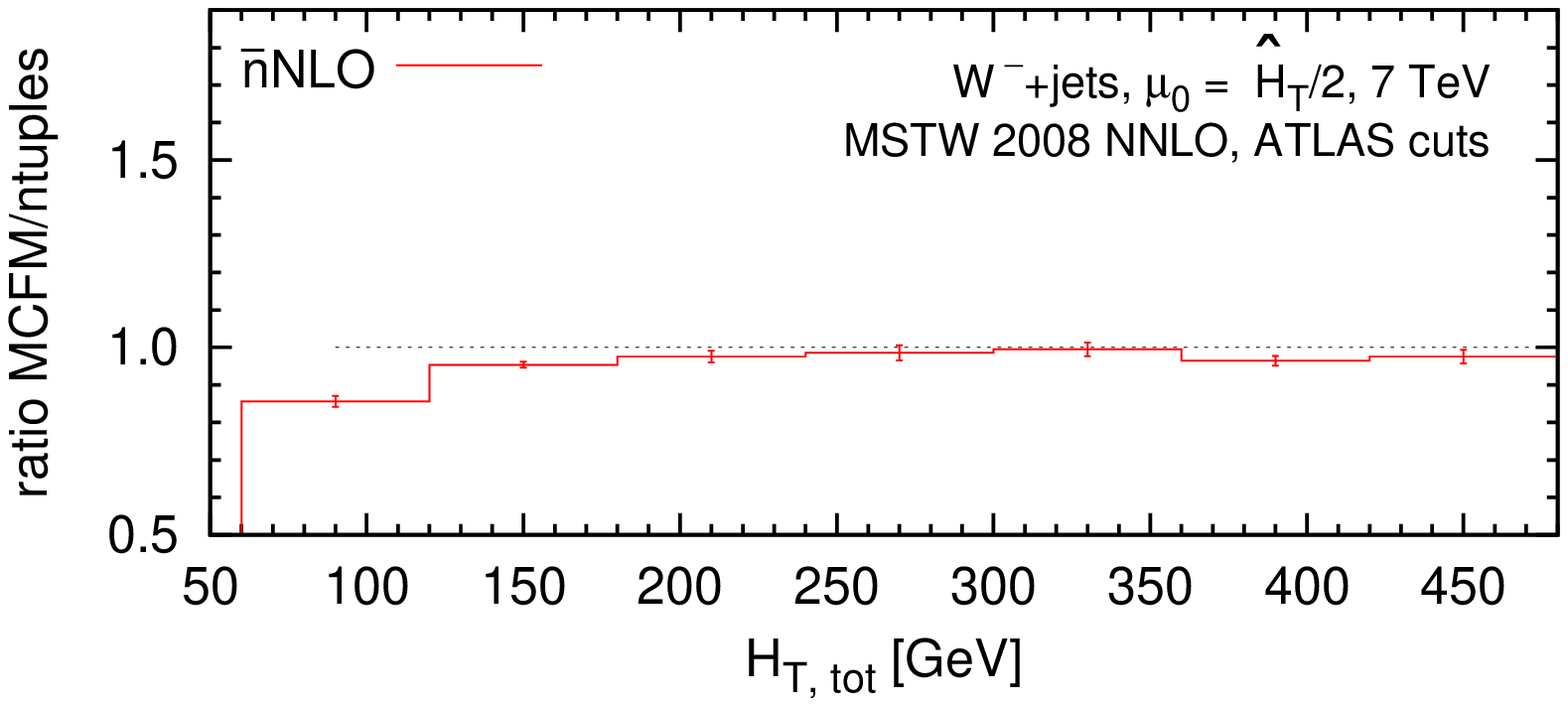}
  \hfill
  \includegraphics[width=0.48\columnwidth]{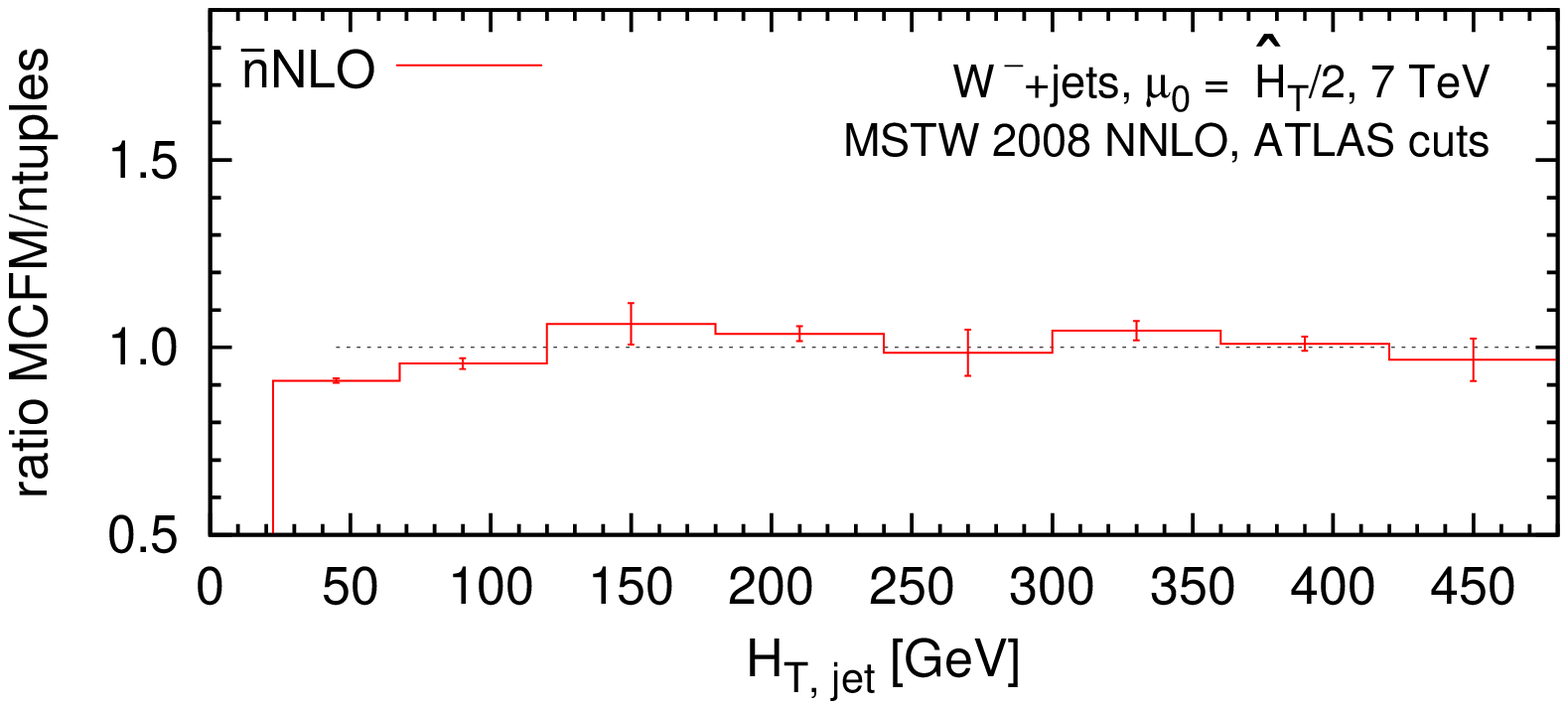}
  \caption{
  Comparison of \nNLO results from LoopSim interfaced with \mcfm~and with
  \bh+\sherpa~ntuples. The bars correspond to statistical errors. The minimal
  jet $p_T = 1 \GeV$  for \mcfm~and 20 \GeV for ntuples. This has an effect on
  the low-$p_T$ bins but the two approaches converge above 100-200 GeV.
  }
  \label{fig:mcfm-vs-ntuples}
\end{figure}
 
The jet radius $R_\LS$ is a parameter of the method. The smaller the value of
$R_\LS$, the more likely the particles are recombined with the beam. Reversely,
the larger $R_\LS$, the more likely the particles are recombined together. The
value of $R_\LS$ is irrelevant for collinear (and, for hardness, also soft)
radiation. It affects only the wide angle (or hard) emissions where the $ij$
mergings compete with the mergings with the beam.
In this study, we shall use $R_\LS = 1$, and we shall vary it by $\pm 0.5$. 
The $R_\LS$ uncertainty will therefore account for the part of the LoopSim
method which is related to attributing the emission sequence and the underlying
hard structure of the events.

The LoopSim method determines exactly the singular (or logarithmic) terms of the
loop diagrams, which, by construction, match precisely the corresponding
singular terms of the real diagrams with one extra parton. Therefore, the \nNLO
result is finite and it differs from the exact NNLO only by the constant terms.
For an observable (A) that receives significant NLO corrections due to new
channels or new topologies, the difference between \nNLO and NNLO, which is
inversely proportional to the K-factor, will be very small
\begin{equation}
  \sigma^{(A)}_\text{\nNLO} = \sigma^{(A)}_{\NNLO} 
  \left( 1 + {\cal O}\left(\frac{\alpha_s^2}{K^{(A)}_{\NNLO}}\right)\right)\,,
  \label{eq:ls-accuracy}
\end{equation}
where $K^{(A)}_\text{NNLO} = \sigma^{(A)}_{\NNLO}/\sigma^{(A)}_{\LO} >
K^{(A)}_\text{NLO} = \sigma^{(A)}_{\NLO}/\sigma^{(A)}_{\LO} \gg 1$.

In order to make the W/Z+jets@\nNLO computation possible, we extended \mcfm~by
adding a reweighting option to allow for efficient generation of tails of
distributions and an option to obtain only the virtual part of the NLO result.
In addition, we added a module that allows to write events in the Les Houches
Event (LHE) format~\cite{Alwall:2006yp}.  Then, we developed an interface
between \mcfm~and LoopSim which uses the LHE for communication between the two
programs.  These extra features of \mcfm~will become part of its new
release~6.7. 
In order to use the results from \bh+\sherpa, we replaced the LHE output
interface to LoopSim with one that accesses the weighted events stored in the
form of \roo~ntuples using the nTupleReader library~\cite{Bern:2013zja}.

The results for W+jets and Z+jets production at \nNLO were obtained both with
LoopSim+\mcfm~and with LoopSim+\roo~ntuples. Although the LoopSim method allows in principle to use event samples with arbitrarily low jet transverse momentum, we need to impose a cut on this quantity for efficiency reasons. 
This technical cut-off on the $p_T$ of jets was set to 1~\GeV for \mcfm~but only
20~\GeV for the ntuples.
In Fig.~\ref{fig:mcfm-vs-ntuples} we show the effect of this different choice on
several distributions. We see that the lack of jets below 20 \GeV in the
\roo~ntuples has an effect only at low $p_T$/$H_T$. Above 100-200 GeV the two
approaches give consistent results at \nNLO.
Since its technical cut-off is lower, we shall only show predictions from
LoopSim interfaced with \mcfm~in the following section. The independence on the
technical cut-off of the prediction at high transverse momenta is encouraging
for the prospect of extending this study to include higher jet multiplicity
samples that are available through \bh+\sherpa~ntuples. 
We have verified that all \nNLO histograms for the observables shown in this
study coincide for the \mcfm~and \bh+\sherpa~ntuples except in immediate
proximity of the generation cut. 

\section{$\mathbf{\nNLO}$ results for W/Z+jets at the LHC at 7 TeV}
\label{sec:Vjetsresults}

In our computation, at all orders, we used the MSTW NNLO 2008
PDFs~\cite{Martin:2009iq}, with $\alpha_s(M_Z)= 0.11707$. Jets were obtained
from clustering final state partons with the anti-$k_t$~\cite{Cacciari:2008gp}
algorithm, using FastJet~\cite{Cacciari:2005hq, FastJet}, with the radius
$R=0.4$. 
The specific cuts for W+jets and Z+jets processes are given in the corresponding
subsections below, and were chosen to match the experimental
analyses~\cite{Aad:2012en} and \cite{Aad:2011qv,Aad:2013ysa}, respectively.
For the central value of the factorization and the renormalization scale, we
chose
\begin{equation}
  \mu_{F,R} = \frac12 \hat H_T = 
  \frac12 \left\{\sum p_{T,\text{partons}}+\sum p_{T,\text{leptons}} \right\}\,,
  \label{eq:mufmur}
\end{equation}
which is a scalar sum of transverse momenta of all particles in the event, i.e.
partons, charged leptons and, if applicable, the neutrino.

The results presented in the following subsections correspond to the parton
level.  In addition, for the plots with comparisons to the experimental data,
and for those plots only, we included the non-perturbative corrections from
hadronization and the underlying event, supplied by
ATLAS~\cite{Aad:2012en,Aad:2013ysa}.
The cross sections correspond to a single lepton channel and, in the case of
W+jets, they include contributions from both W$^+$ and W$^-$.

\subsection{W+jets}
\label{sec:wjets}

The analysis is performed with the following cuts~\cite{Aad:2012en}:
The charged leptons are required to have $p_{T,\ell}\ge 20 \GeV$ and
$|y_{\ell}|\le 2.5$. The missing transverse energy must be above
$E_{T,\text{miss}} > 25 \GeV$.  The transverse mass of the W, defined as
$m_{T,W} = \sqrt{2p_{T,\ell}\,p_{T,\nu}(1-\cos(\phi_\ell-\phi_\nu))}$, is
required to be greater than $40 \GeV$. Only events with jets with $p_{T,\jet} >
30 \GeV$ and $|y_{\jet}| < 4.4$ are accepted. In addition, if the distance
between a jet and a lepton $\Delta R(\ell,\jet)$ is smaller than 0.5, this jet
is removed from the list of the jets, but the event is kept, as long as the
other requirements are met.
\begin{figure}[p]
  \includegraphics[width=0.48\columnwidth]{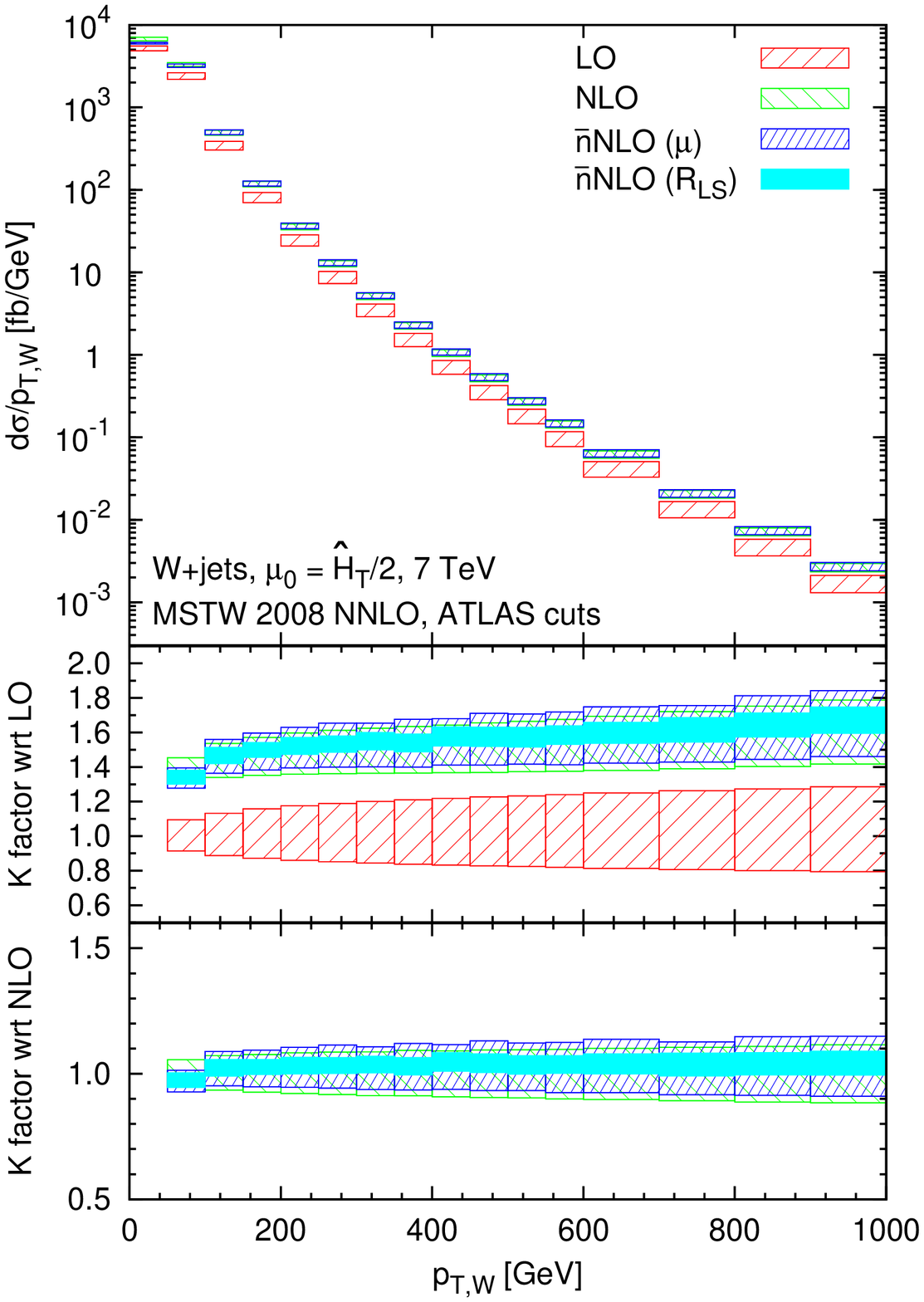}
  \hfill
  \includegraphics[width=0.48\columnwidth]{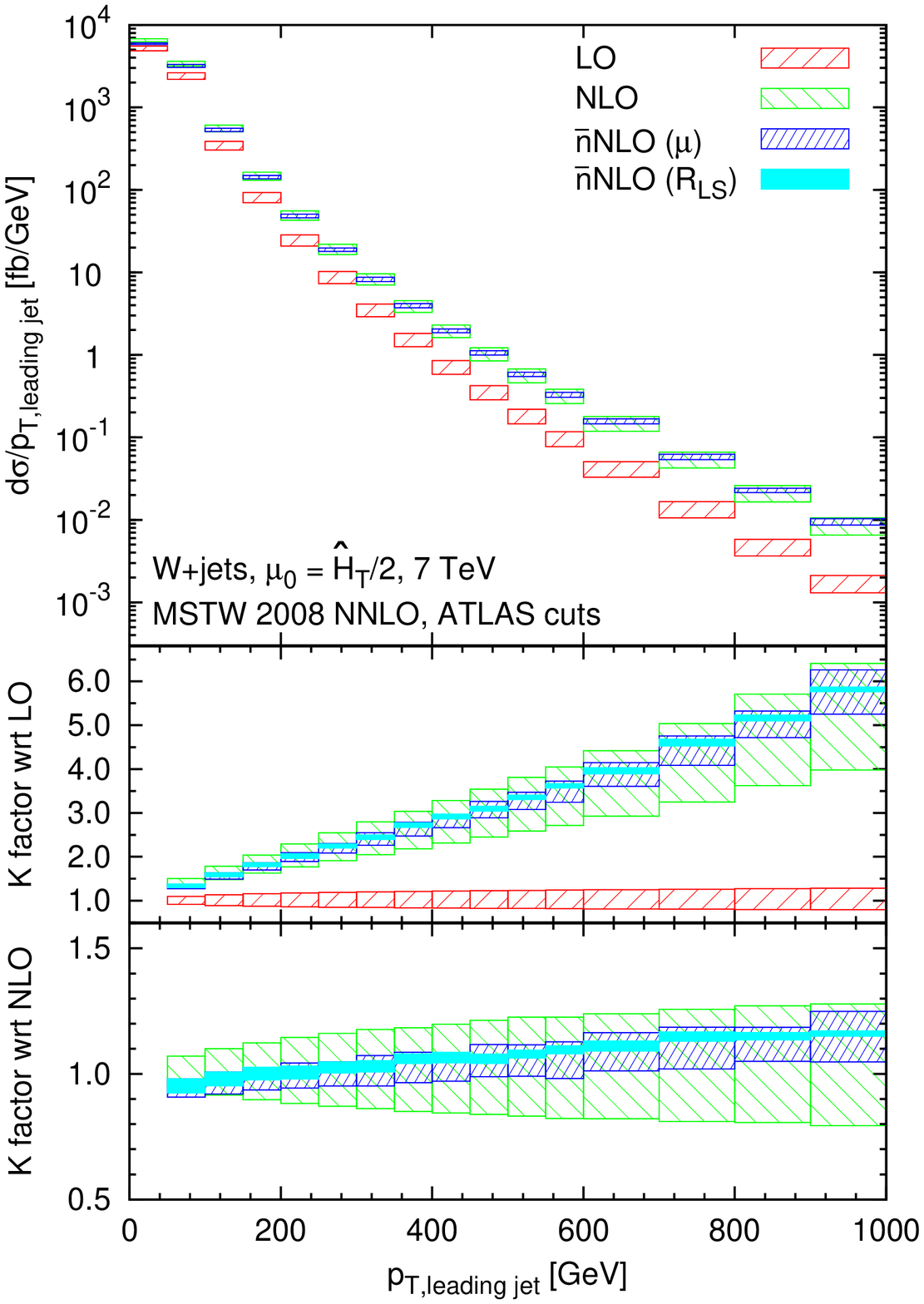}
  \includegraphics[width=0.48\columnwidth]{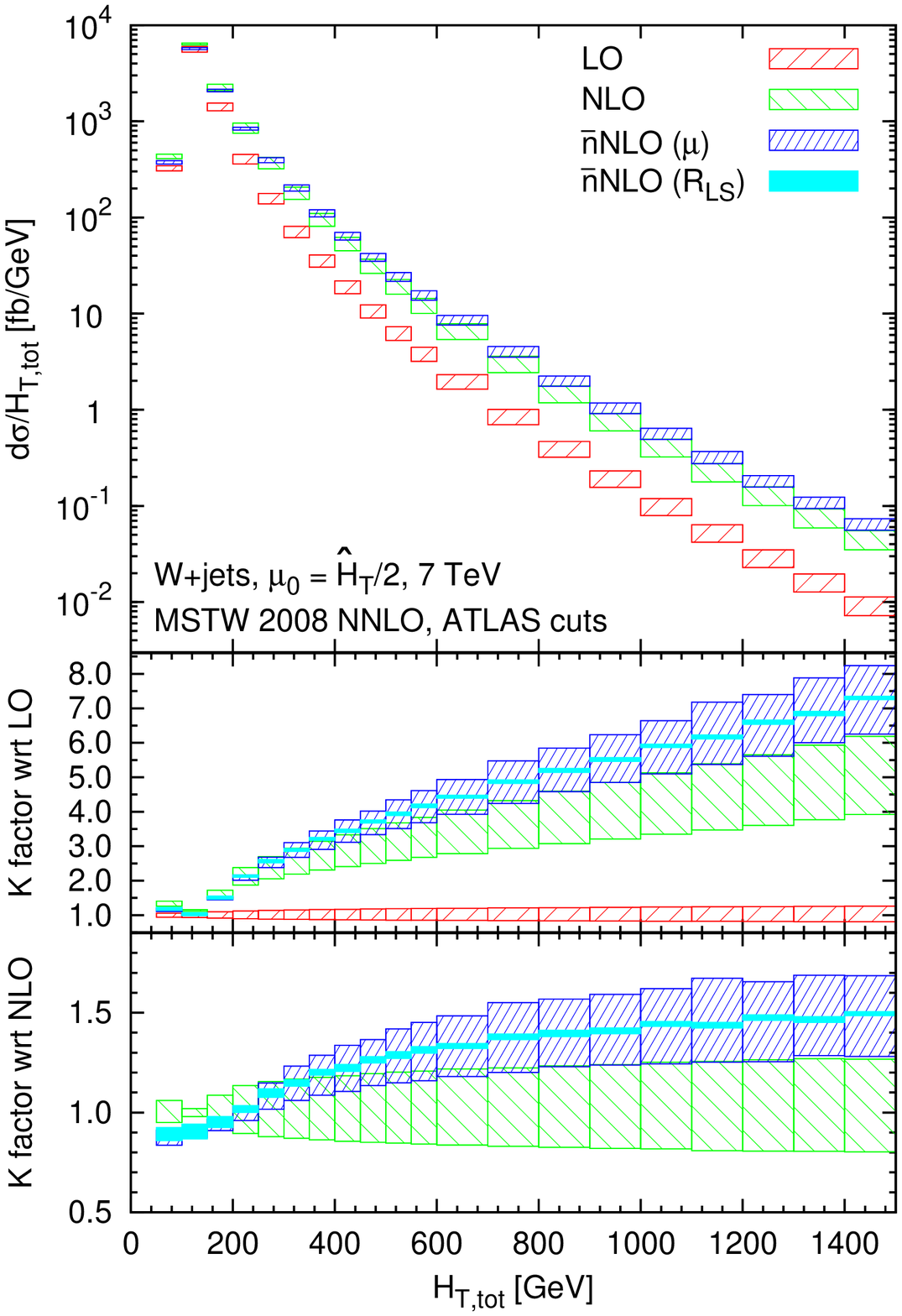}
  \hfill
  \includegraphics[width=0.48\columnwidth]{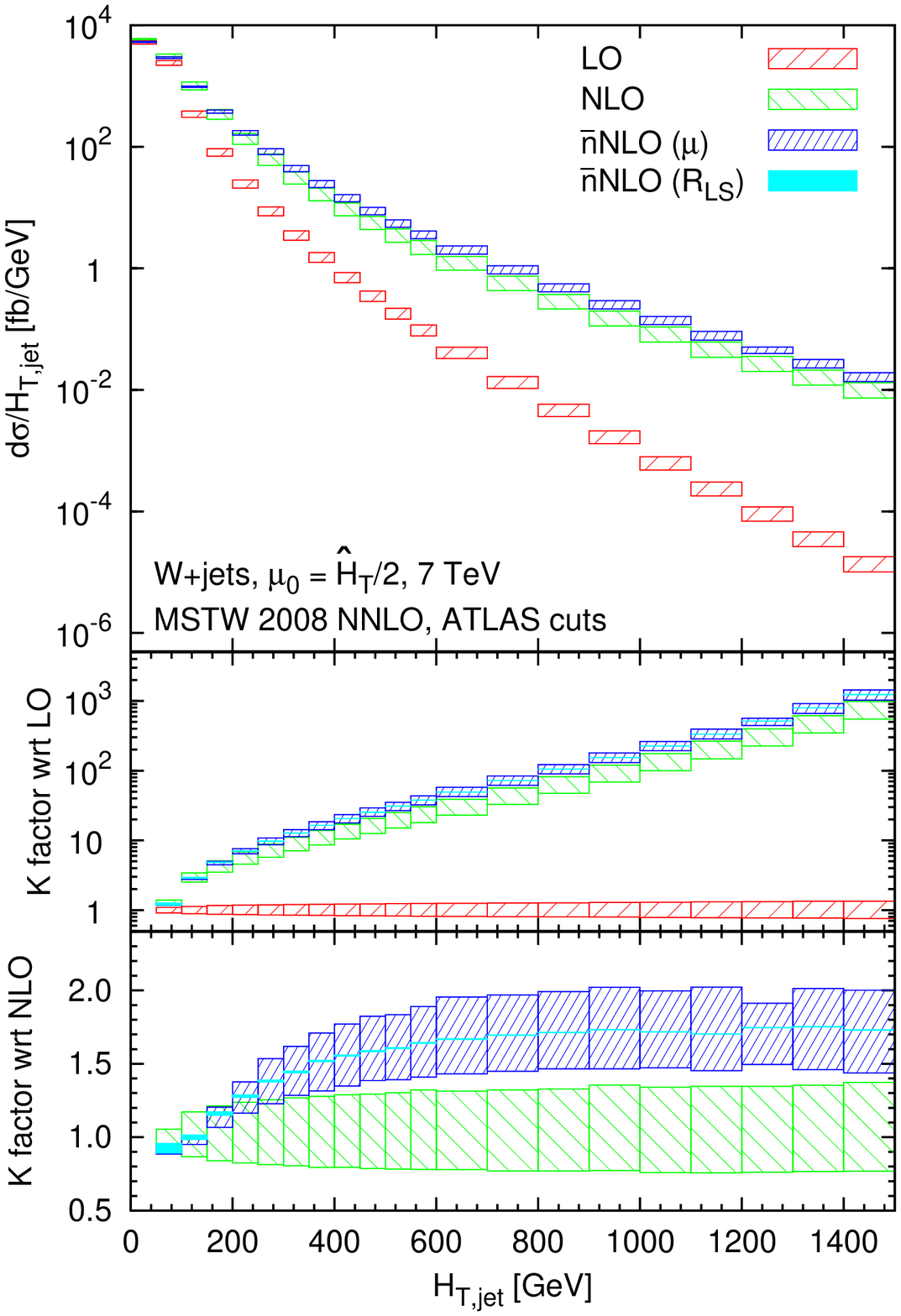}
  \caption{
  Differential cross sections and K-factors for $p_{T,W}$, $p_{T,\text{leading
  jet}}$, $H_{T,\jet}$ and  $H_{T,\tot}$ at parton level at LO, NLO and \nNLO.
  The bands correspond to varying $\mu_F=\mu_R$ by factors 1/2 and 2
  around the central value from Eq.~(\ref{eq:mufmur}) or to changing $R_\LS$
  to 0.5 and 1.5.
  The distributions in this and the following figures are sums of contributions
  from $W^+$ and $W^-$ and correspond to a single lepton decay channel $W\to
  \ell \nu$.
  }
  \label{fig:wjets-allobs1}
\end{figure}
\begin{figure}[t]
  \includegraphics[width=0.48\columnwidth]{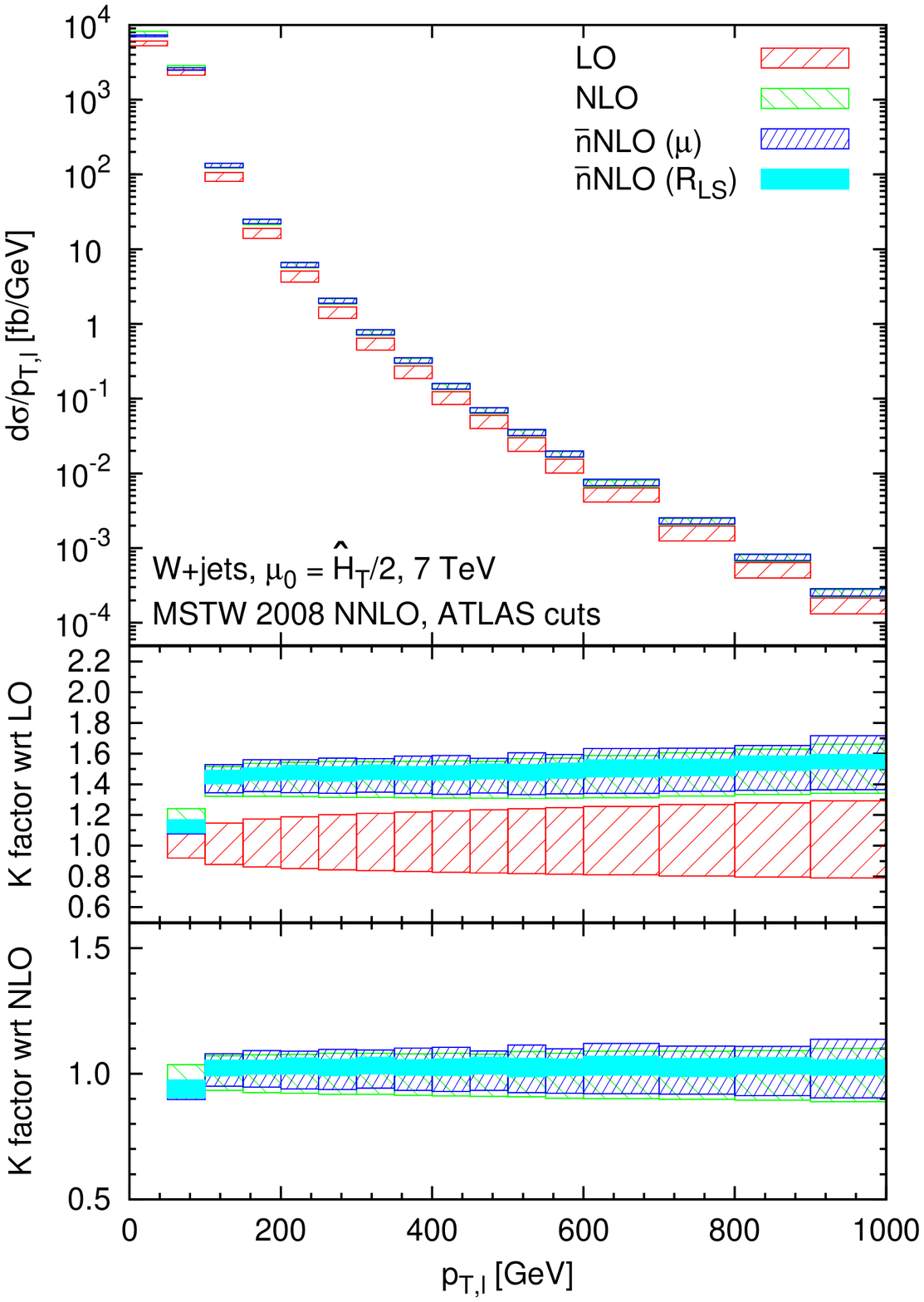}
  \hfill
  \includegraphics[width=0.48\columnwidth]{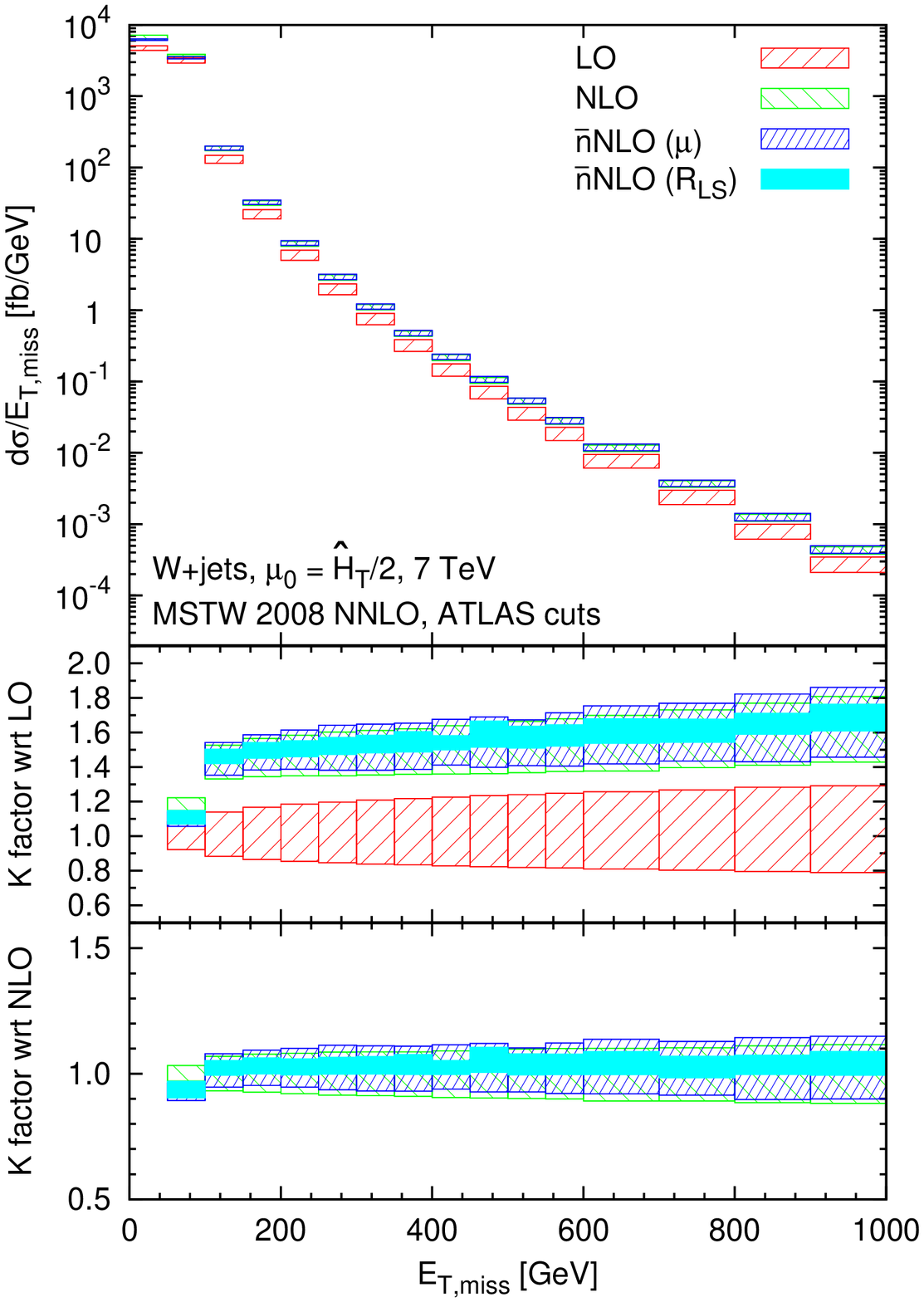}
  \caption{
  Differential cross sections and K-factors for $p_T$ of the lepton from the
  decay of the W boson and missing energy $E_{T,\miss}$ at parton level at LO,
  NLO and \nNLO.
  The bands correspond to varying $\mu_F=\mu_R$ by factors 1/2 and 2 around the
  central value from Eq.~(\ref{eq:mufmur}) or changing $R_\LS$ to 0.5 and
  1.5.
  }
  \label{fig:wjets-allobs2}
\end{figure}

In Figs.~\ref{fig:wjets-allobs1} and \ref{fig:wjets-allobs2}, we present the
differential distributions for the following observables:
\begin{itemize}
\item $p_{T,V}$: transverse momentum of the vector boson,
\item $p_{T,\text{leading jet}}$: transverse momentum of the hardest jet,
\item $H_{T,\rm{tot}}$: sum of the transverse momenta of the leptons (including
the neutrino) and jets passing the jet cuts,
\item $H_{T,\rm{jets}}$: sum of the transverse momenta of the jets passing the jet cuts.
\end{itemize} 
The predictions are at the pure parton level for $\sqrt{s} = 7 \TeV$.  In
each plot, the top panel shows the distributions at LO, NLO and \nNLO whereas
the middle and the bottom ones depict the K-factors with respect to LO and NLO.

The distribution of the $p_T$ of the W boson, shown in
Fig.~\ref{fig:wjets-allobs1}, is fairly insensitive to new topologies
appearing at NLO, and therefore does not exhibit a large NLO/LO K-factor and
does not receive significant corrections at \nNLO.
In contrast, the distribution of the $p_T$ of the leading jet from
Fig.~\ref{fig:wjets-allobs1} shows a large K-factor at NLO, which grows with
$p_T$ and which arises due to new topologies with soft and collinear W boson.
This observable is therefore very suitable for LoopSim predictions at \nNLO, cf.
Eq.(\ref{eq:ls-accuracy}).
Indeed, we see a substantial (almost 70\% at high $p_T$) reduction of the scale
uncertainty at \nNLO, while the result stays within the NLO band. That indicates
that the distribution of the leading jet $p_T$ comes under control at \nNLO,
which matches the expectations, as no new channel or topologies appear at this
order. We also note that the uncertainty due to the $R_\LS$ variation is smaller
than the scale uncertainty and it decreases with increasing $p_T$.

As mentioned in Sec.~\ref{sec:details}, and explained in more detail
in~\cite{Rubin:2010xp}, the LoopSim method is expected to give accurate
predictions in the region in which a substantial part of the NLO/LO K-factor
comes from new channels or new topologies, as opposed to the constant loop
terms.
As shown in the middle panel of the $p_{T,W}$ distribution in
Fig.~\ref{fig:wjets-allobs1}, the K-factor, which comes predominantly from
genuine loop effects of the LO-type topology diagrams, is around 1.5. Those
LO-type topologies should still give significant contributions to the low $p_T$
region of the leading jet $p_T$ distributions and we expect that new topologies
start giving the dominant corrections around the K-factor $\sim 2.5$, which
corresponds to $p_{t,\text{leading jet}} \sim 400 \GeV$. Therefore, the \nNLO
result from LoopSim is expected to be accurate above that value of $p_T$.

Fig.~\ref{fig:wjets-allobs1} shows also the $H_{T,\tot}$ and $H_{T,\jet}$
distributions. We see that both receive large corrections at \nNLO reaching
above 50\% at 1.5 TeV, and lying outside the NLO bands. The reduction of the
scale uncertainty at \nNLO is only minor for these observables, however, do to
the 50\% increase, the relative scale uncertainty is smaller than that of the
NLO result.  We also see that the $R_\LS$ uncertainty is negligible.
The large \nNLO corrections to $H_{T,\tot}$ and $H_{T,\jet}$ come from the third
jet, present due to initial state radiation. This jet's $p_T$ adds a small
contribution to $H_{T}$ but, because the spectrum is steeply falling, the
enhancement in the distribution is substantial. 

\begin{figure}[t]
  \begin{center}
    \includegraphics[width=0.9\columnwidth]{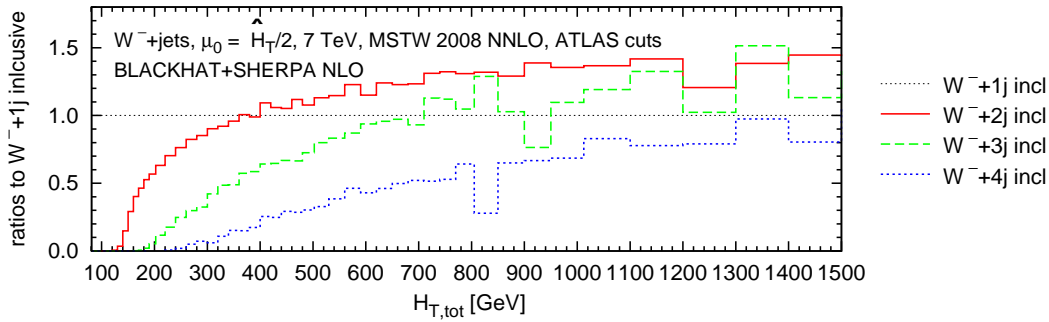}
  \end{center}
  \caption{
  NLO predictions for the $H_{T,\tot}$ distribution for different inclusive jet multiplicities. All curves are normalised to the inclusive 1-jet prediction.
  }
  \label{fig:multip-contr}
\end{figure}
 
The importance of including higher multiplicities for the description of
$H_{T,\tot}$ is illustrated in Fig.~\ref{fig:multip-contr}. One sees that as the
value of $H_{T,\tot}$ increases, the NLO prediction for the higher multiplicity
increases relative to the inclusive 1-jet contribution, the latter being
overtaken by the 2-jet prediction at the $H_{T,\tot}$ value of $350\;\rm{GeV}$
and by the 3-jet contribution at the value of $650\;\rm{GeV}$. This fact
demonstrates the need to include higher order effects to describe the
$H_{T,\tot}$ distribution. In our calculation we include the 1 and 2-jet
contribution at NLO and the 3-jet contribution at LO.  We
expect the 3 and 4-jet topologies at NLO to play a role above 600 GeV. It would
therefore be interesting to study the inclusion of these higher multiplicities,
along with a simulated estimate of the corresponding loop effects, using the
LoopSim method.
 
Finally, in Fig.~\ref{fig:wjets-allobs2}, we present the
results for the $p_T$ of the lepton and missing transverse energy. Both of these
distributions, just like the $p_{T,W}$ from Fig.~\ref{fig:wjets-allobs1} are
not sensitive to new topologies appearing at NLO and therefore they are well
behaved already at that order. Consequently, the \nNLO result from LoopSim does
not lead to any significant difference with respect to NLO.

\begin{figure}[t]
  \includegraphics[width=0.5\columnwidth]{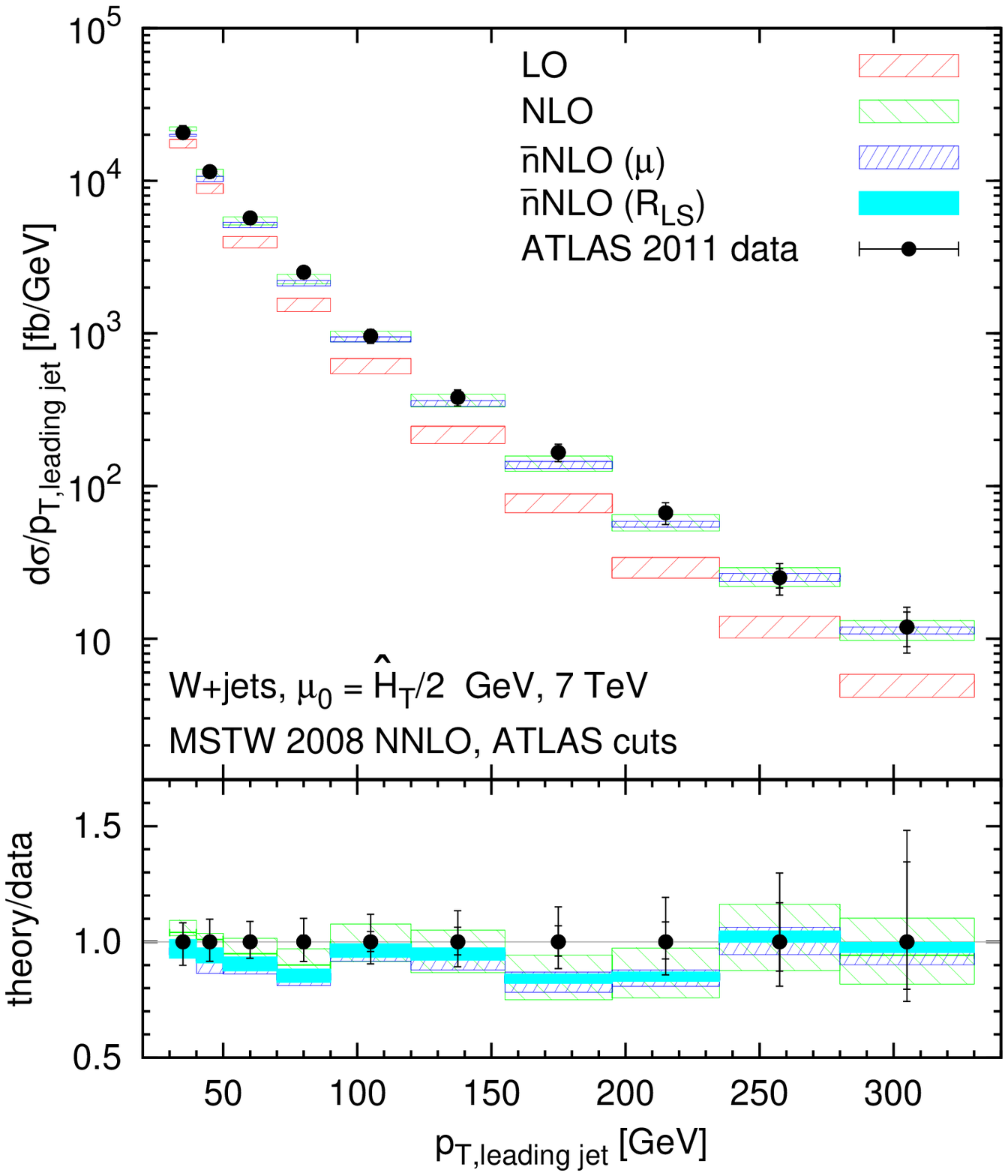}
  \hfill
  \includegraphics[width=0.5\columnwidth]{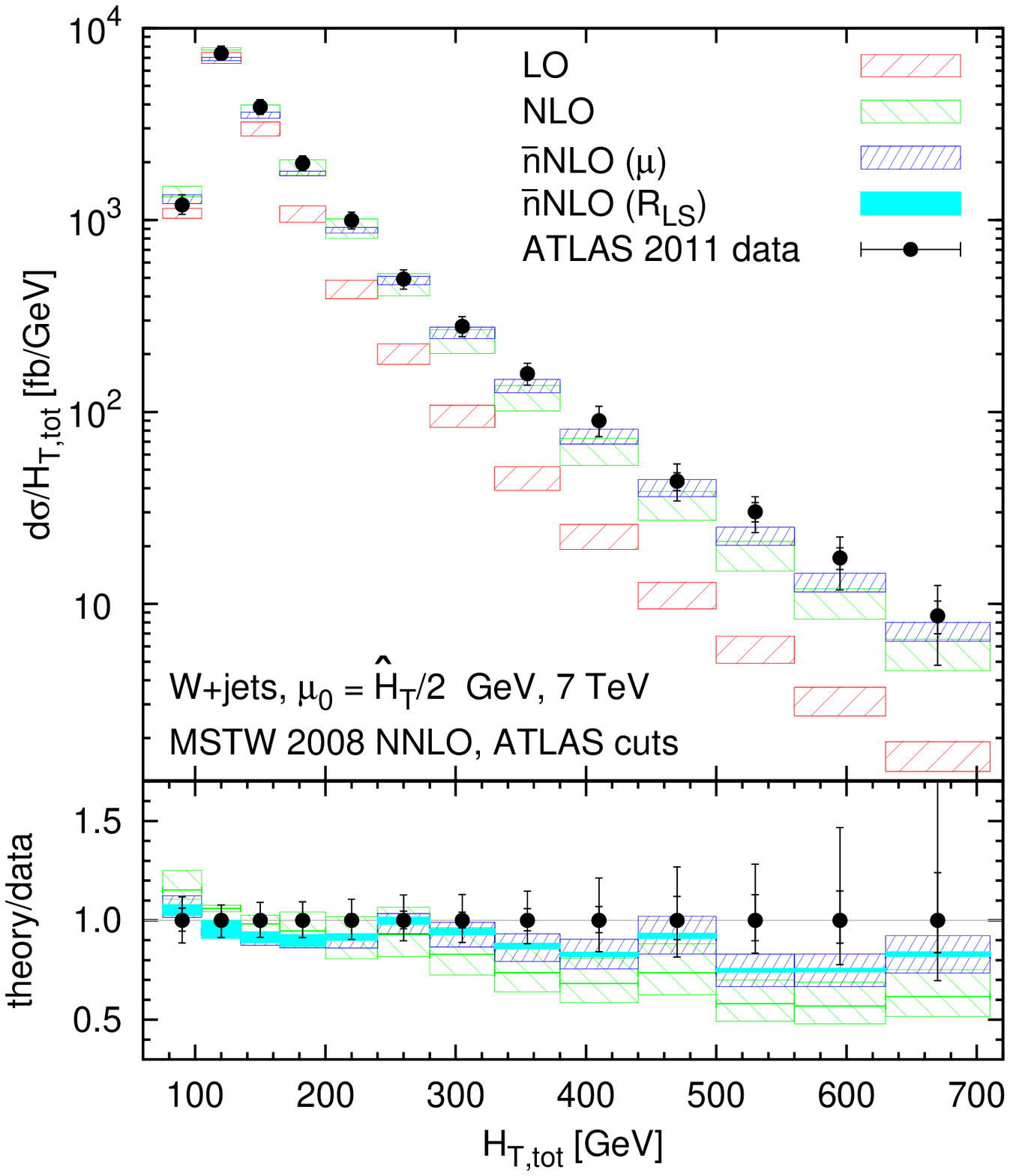}
  \caption{
  Differential cross sections for W+jets production at LO, NLO and \nNLO as
  functions of the $p_T$ of the hardest jet (left) and the scalar sum of $p_T$s
  of all particles (right).
  The theoretical results are corrected for hadronization and UE effects and are
  compared to the ATLAS 2010 7~TeV data of $36\, \pb^{-1}$~\cite{Aad:2012en},
  with the lower panel showing the ratio of the two. The inner and outer bars
  on the data points give statistical and total errors, respectively.
  The bands correspond to varying $\mu_F=\mu_R$ by factors 1/2 and 2 around the
  central value from Eq.~(\ref{eq:mufmur}). The cyan solid bands give the
  uncertainty related to the $R_\LS$ parameter changed to 0.5 and 1.5.  
  }
  \label{fig:wjets-data}
\end{figure}

In Fig.~\ref{fig:wjets-data}, we compare our predictions to the existing,
$\sqrt{s} = 7 \TeV$, data from ATLAS~\cite{Aad:2012en}.
Fig.~\ref{fig:wjets-data} (left) shows the differential distributions for the
transverse momentum of the hardest jet and Fig.~\ref{fig:wjets-data} (right),
the scalar sum of the traverse momenta of jets, leptons and missing energy,
$H_{T,\tot}$.  The inner and outer bars on the data points correspond to
statistical and total error, respectively.
The theoretical predictions computed at LO, NLO and \nNLO orders
in QCD were corrected for hadronization and UE effects using the coefficients
determined in~\cite{Aad:2012en}.
 
The \nNLO prediction for the distribution of the $p_T$ of the leading jet, 
stays within the NLO band and its scale uncertainty is significantly reduced
with respect to NLO. At low $p_T$, it touches the lower edge of the NLO band
but, as discussed above,  this region is potentially sensitive to the
genuine constant terms of the loop diagrams, which are not guaranteed to be
determined precisely by the LoopSim method.
In the case of $H_{T,\tot}$, the \nNLO result goes
beyond the NLO uncertainty band for $H_{T,\tot}> 300 \GeV$ and the corrections
are up to 30\% with respect to NLO. The $R_\LS$ uncertainty becomes very small
above 300 GeV.  
As we see in Fig.~\ref{fig:wjets-data} (right), the \nNLO result, by including
configurations with three partons in the final state, describes the $H_{T,\tot}$
data better than the inclusive W+1 jet NLO prediction. It would be interesting
to establish whether inclusion of the \nnNLO corrections (\ie~exact 1-loop and
simulated 2 and 3-loops) would further improve the agreement with the data. We
leave this question for future work.

\subsection{Z+jets}
\label{sec:zjets}

\begin{figure}[p]
  \includegraphics[width=0.48\columnwidth]{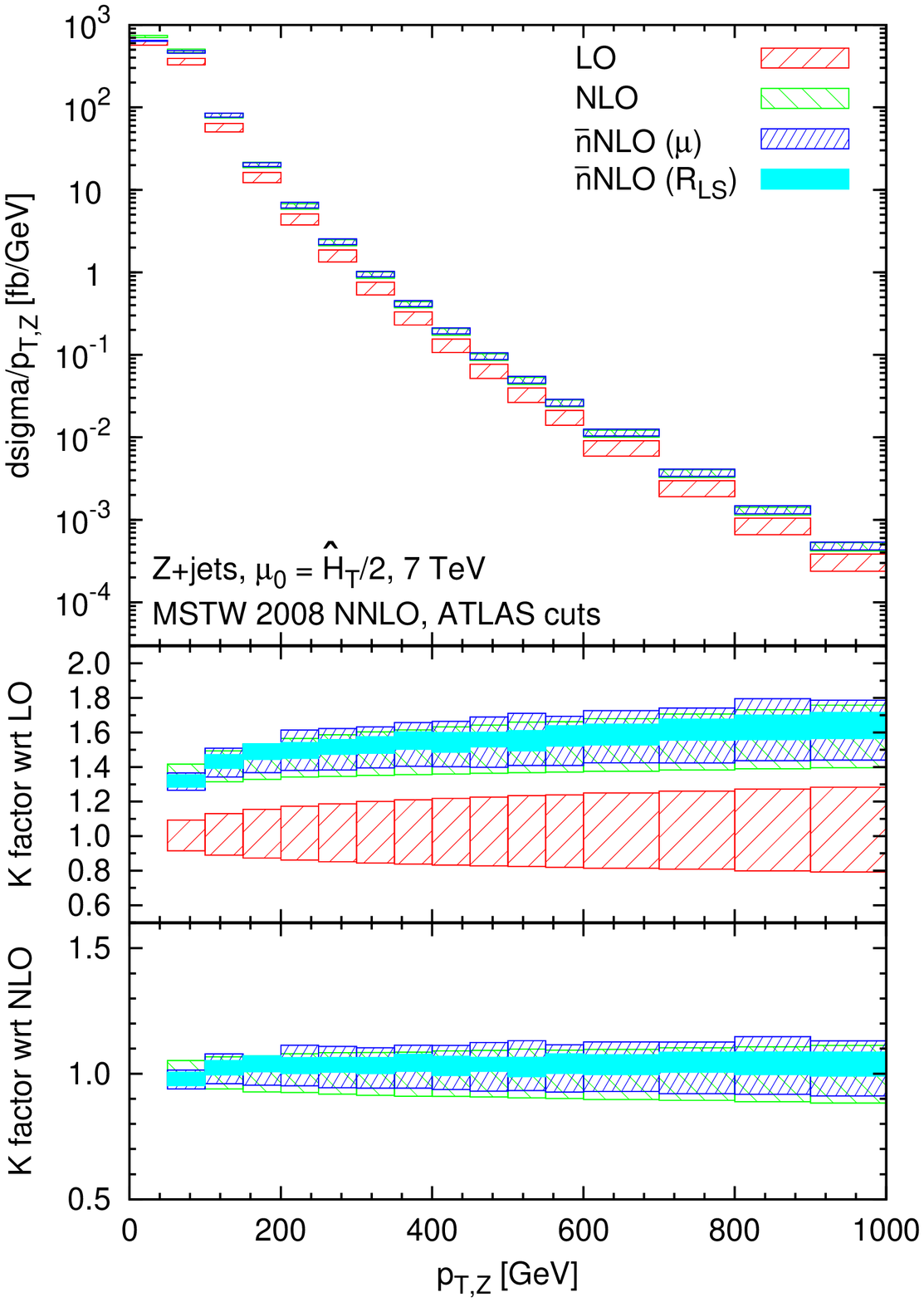}
  \hfill
  \includegraphics[width=0.48\columnwidth]{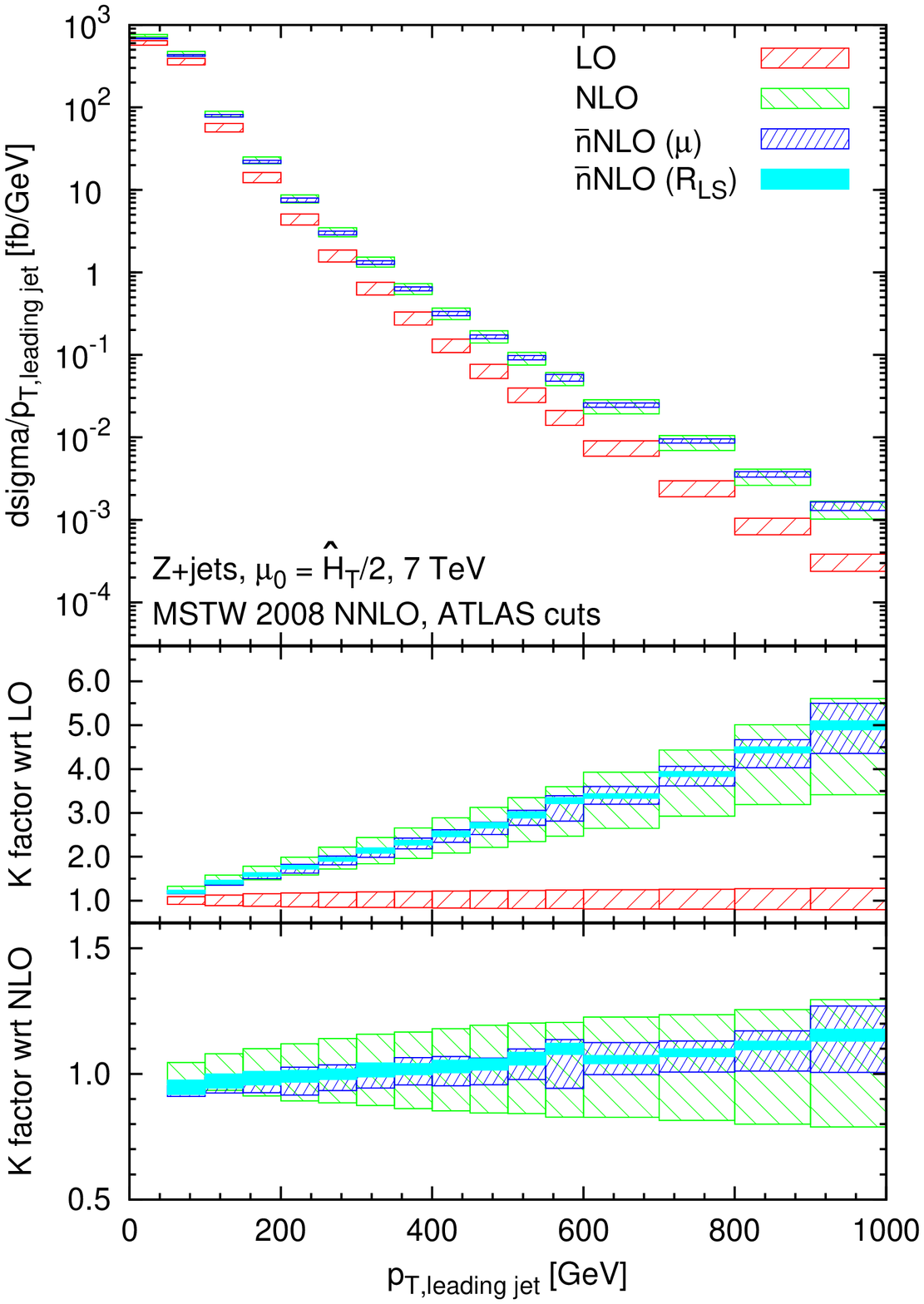}
  \includegraphics[width=0.48\columnwidth]{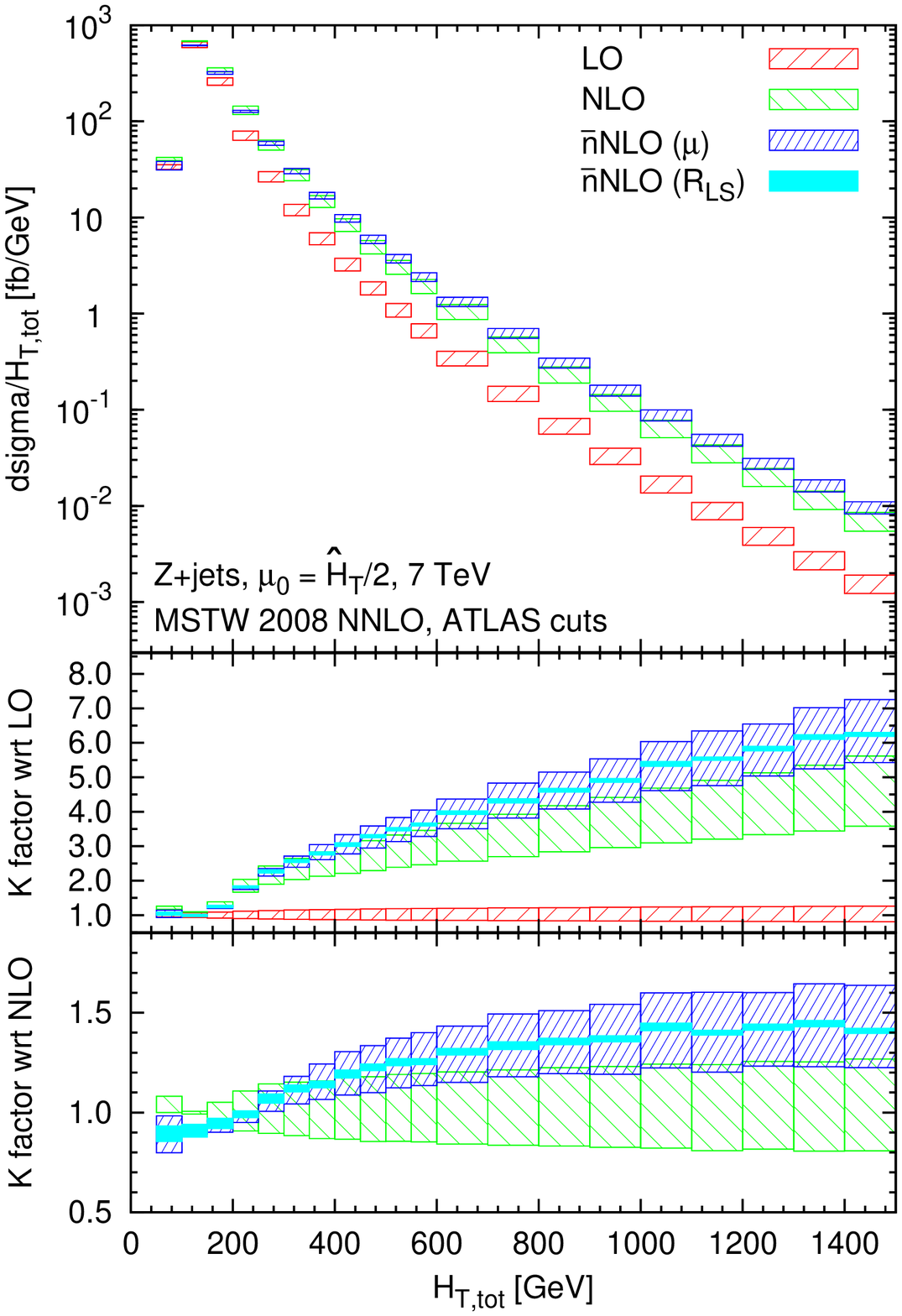}
  \hfill
  \includegraphics[width=0.48\columnwidth]{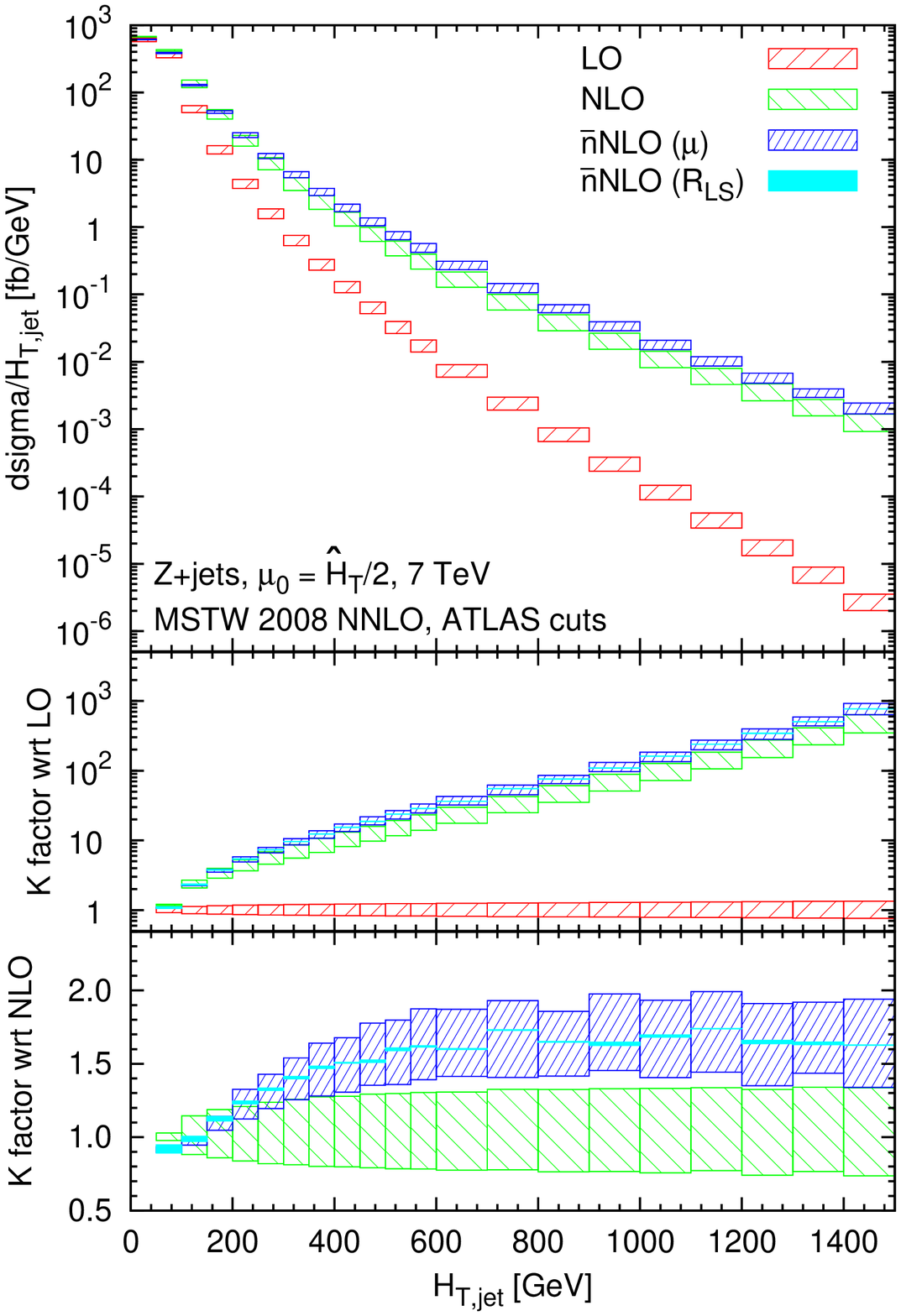}
  \caption{
  Differential cross sections and K-factors for $p_{T,Z}$, $p_{T,\text{leading
  jet}}$, $H_{T,\jet}$ and  $H_{T,\tot}$ at parton level at LO, NLO and \nNLO.
  The bands correspond to varying $\mu_F=\mu_R$ by factors 1/2 and 2
  around the central value from Eq.~(\ref{eq:mufmur}) or changing $R_\LS$
  to 0.5 and 1.5.
  \label{fig:zjets-allobs1}
  }
\end{figure}

This process involves the production of a Z boson with one or more jets,
including the effects of interference with virtual photon. The Z boson
subsequently decays into a pair of electrons or muons.

The analysis is performed with the following cuts~\cite{Aad:2011qv,Aad:2013ysa}:
The charged leptons are required to have $p_{T,\ell}\ge 20 \GeV$ and
$|y_{\ell}|\le 2.5$ and the dilepton mass must lie in the window $66 <
m_{\ell\ell} < 116 \GeV$.
The jets are required to be sufficiently hard and central with $p_{T,\jet} > 30
\GeV$ and $|y_{\jet}| < 4.4$ and, similarly to the W+jets case, they are removed
from an event if $\Delta R(\ell,\jet)~>~0.5$.
\begin{figure}[p]
  \includegraphics[width=0.48\columnwidth]{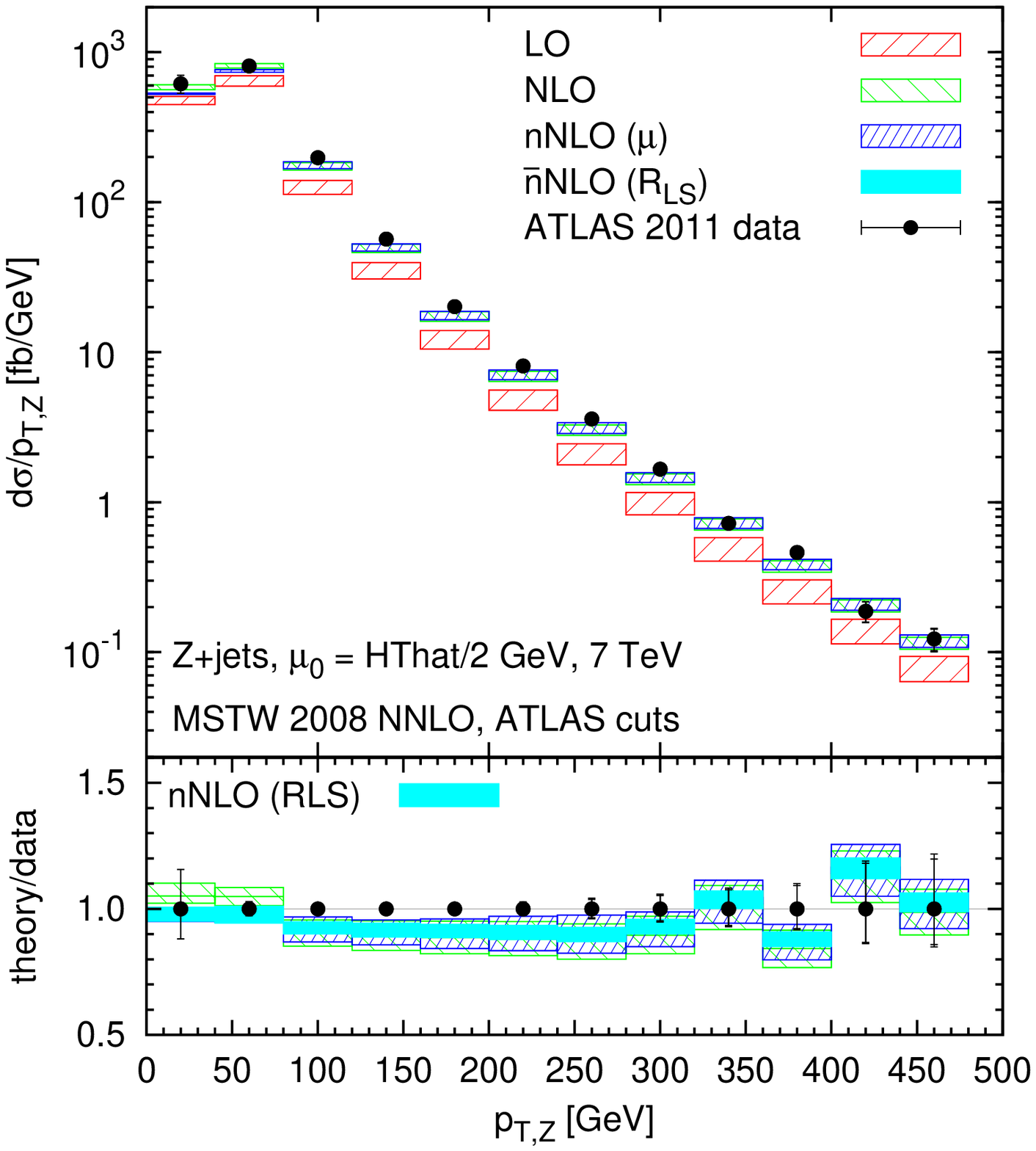}
  \hfill
  \includegraphics[width=0.48\columnwidth]{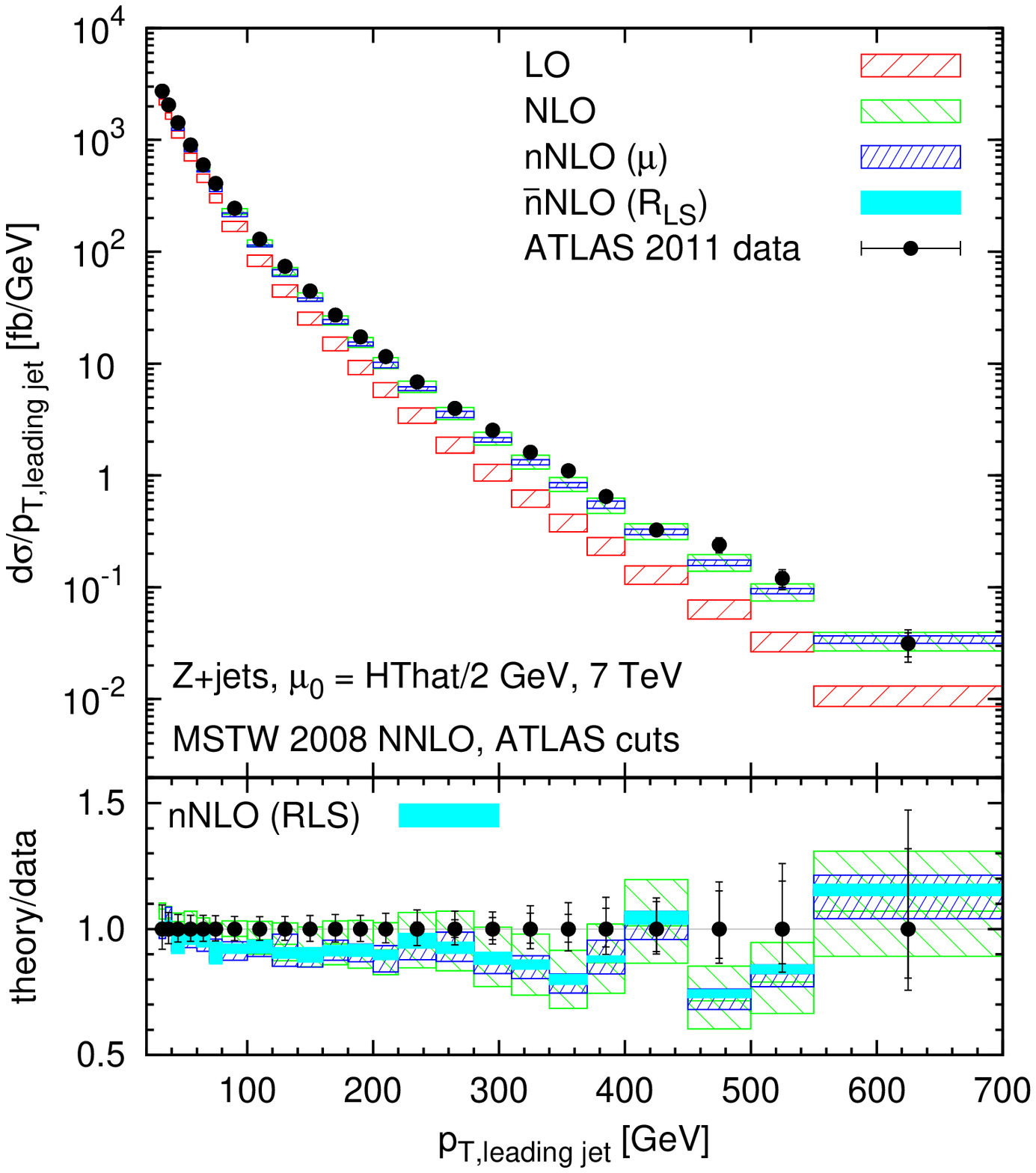}
  \includegraphics[width=0.48\columnwidth]{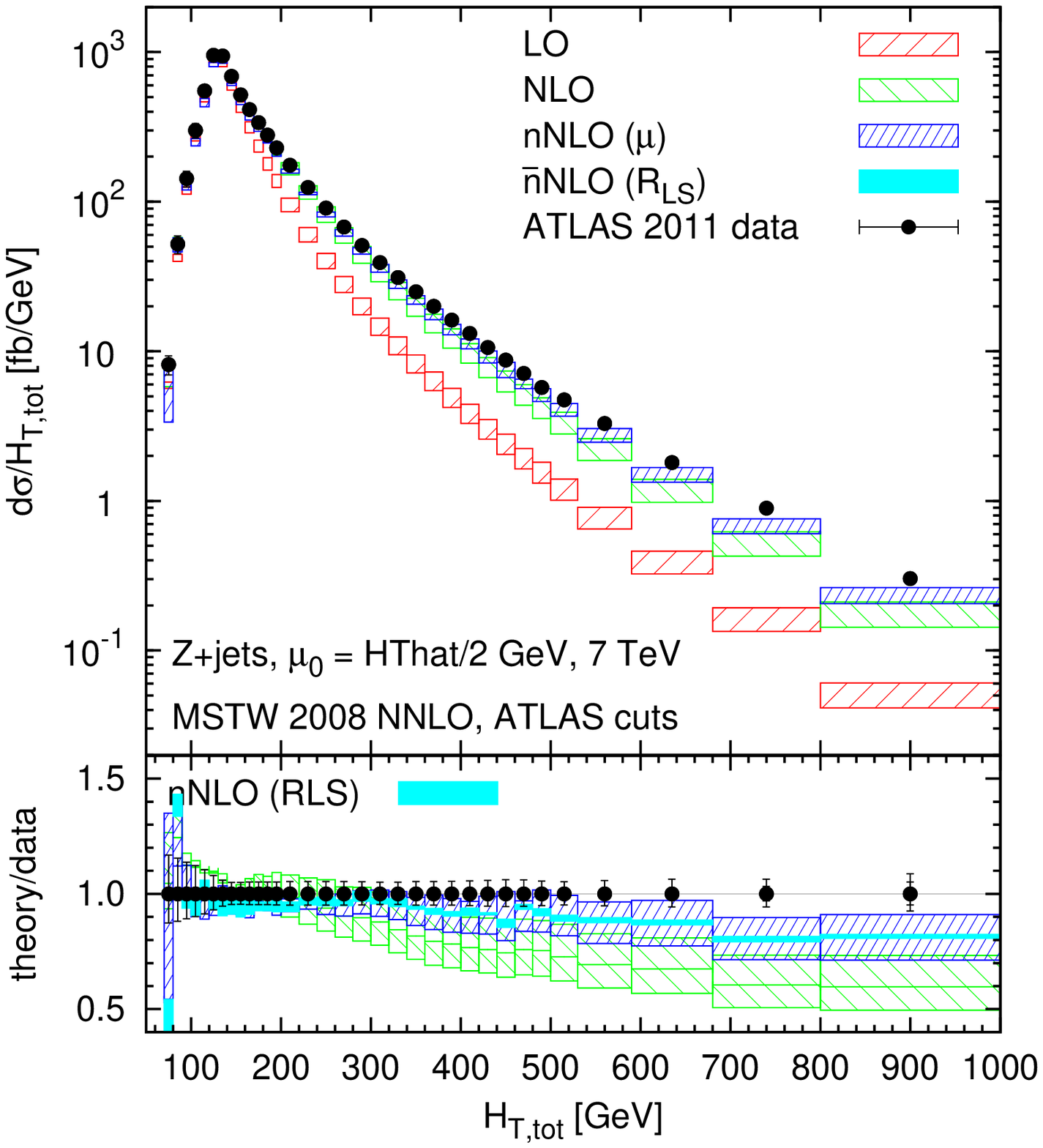}
  \hfill
  \includegraphics[width=0.48\columnwidth]{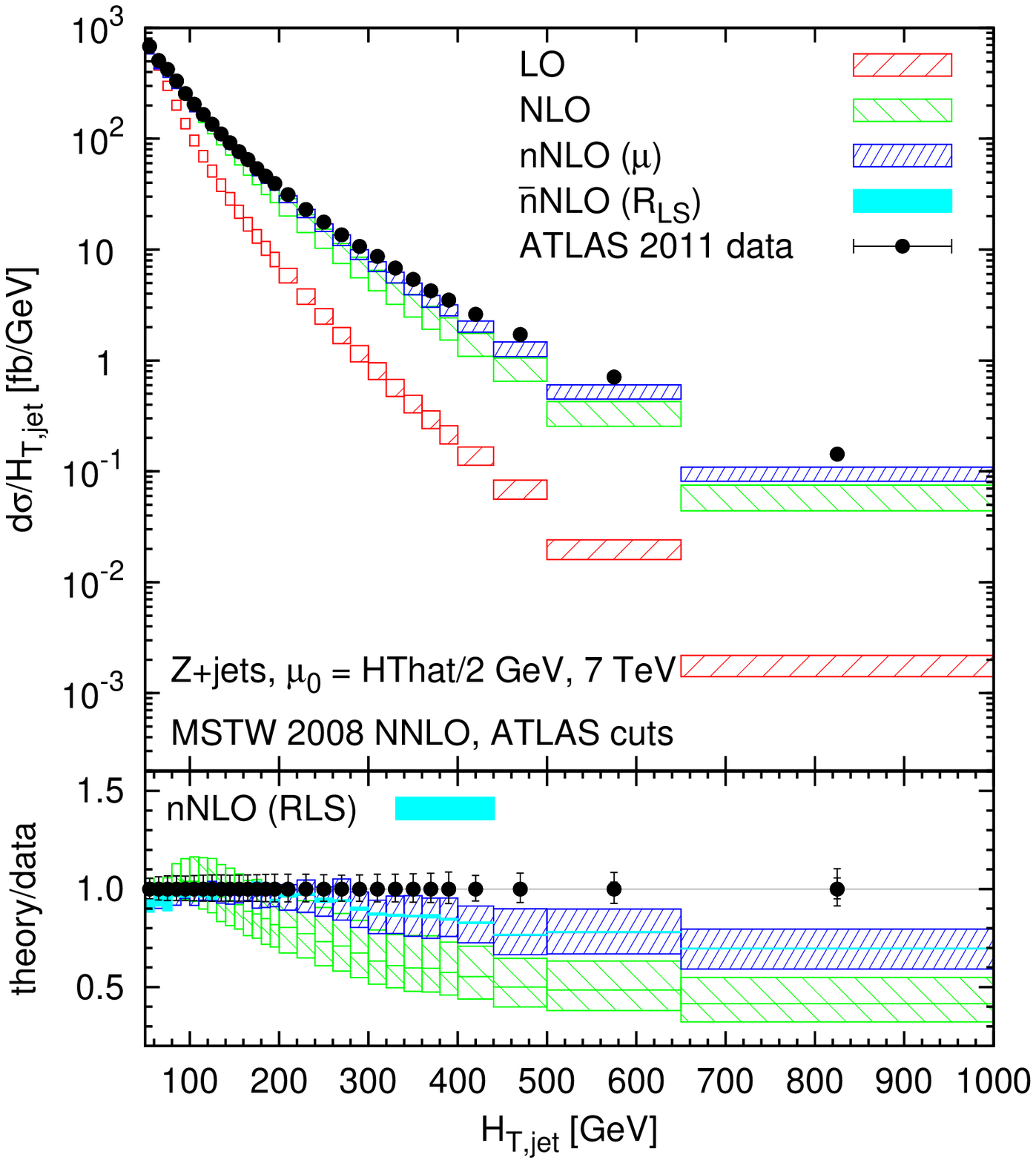}
  \caption{
  Differential cross sections for Z+jets production at LO, NLO and \nNLO as
  functions of the $p_T$ of the Z boson, $p_T$ of the hardest jet, the scalar
  sum of the transverse momenta of all particles $H_{T,\tot}$ and the scalar sum
  of the transverse momenta of jets $H_{T,\jet}$.
  The theoretical results are corrected for hadronization, UE and QED effects
  and are compared to the ATLAS 2011 7 TeV data of $4.6\,
  \fb^{-1}$~\cite{Aad:2013ysa}. The lower panel shows predictions
  normalised to data.
  The bands correspond to varying factorization and renormalization scales
  $\mu_F=\mu_R$ by factors 1/2 and 2 around the central value from
  Eq.~(\ref{eq:mufmur}). The cyan solid bands give the uncertainty related to
  the $R_\LS$ parameter changed to 0.5 and 1.5.  The distributions
  correspond to a single lepton decay channel $Z\to \ell \ell$.
  }
  \label{fig:zjets-data}
\end{figure}

In Fig.~\ref{fig:zjets-allobs1}, we show several distributions relevant for this
process.  All details are the same as in the W+jets case discussed above.  Also,
the qualitative behaviour is similar. Thus, the distribution of the $p_T$ of the
Z boson does not receive any significant corrections at \nNLO, which is related
to the small NLO/LO K-factor for this observable and to its low sensitivity to
new topologies appearing at NLO.~\footnote{This does not imply that
there are no significant contributions coming from constant terms of the 2-loop
diagrams, as we discuss further in this section.}
On the other hand, the distribution of the $p_T$ of the leading jet is
significantly improved at \nNLO and comes under control at this order. As
discussed in the previous subsection, LoopSim becomes powerful in predicting
this distribution above $p_{T,\text{leading jet}} \sim 300-400 \GeV$.
The $H_{T,\tot}$ and $H_{T,\jet}$ distributions receive large, up to 50\%
corrections at \nNLO and this, together with the small reduction of scale
uncertainties, indicates that they are still not converging at this order due to
lacking multiplicities.
We also computed the distributions of the $p_T$ of the leading and $p_T$ of the
trailing lepton (results not displayed) and, as in the case of the $p_{T,\ell}$
and $E_{T,\miss}$ of the W+jets process, they do not show an improvement at
\nNLO.

In Fig.~\ref{fig:zjets-data} we compare our predictions with the ATLAS results
obtained at $\sqrt{s} = 7 \TeV$~\cite{Aad:2013ysa}.  The distributions
correspond to a single lepton channel.  The inner and outer bars on the data
points correspond to statistical and total error, respectively.
The theoretical results at LO, NLO and \nNLO were corrected for hadronization
and UE effects, as well as for QED effects, following the procedure used in
~\cite{Aad:2013ysa}.
Moreover, in the theory/data ratio, the experimental results were normalized
to the inclusive Z cross section, measured in the same fiducial volume whereas
the theory results were normalized using an analogous result from
\bh+\sherpa.

As expected, the $p_{T,Z}$ distribution does not receive an improvement at \nNLO
and the offset in the theory/data ratio in the region around $100-300 \GeV$,
observed at NLO, stays also at \nNLO. This could, in principle, point to
non-negligible constant terms of the 2-loop diagrams.
The distribution of the leading jet $p_T$ shows a significant reduction of the
scale uncertainty at \nNLO and a small dependence on $R_\LS$ at higher $p_T$
values.  

The $H_{T,\jet}$ and  $H_{T,\tot}$ results at \nNLO compare significantly better
to the experimental data than plain NLO. They quickly go outside of the NLO
bands and their scale uncertainty is only moderately reduced.


\subsection{Ratios}
\label{sec:ratios}

\begin{figure}[t]
  \includegraphics[width=0.5\columnwidth]{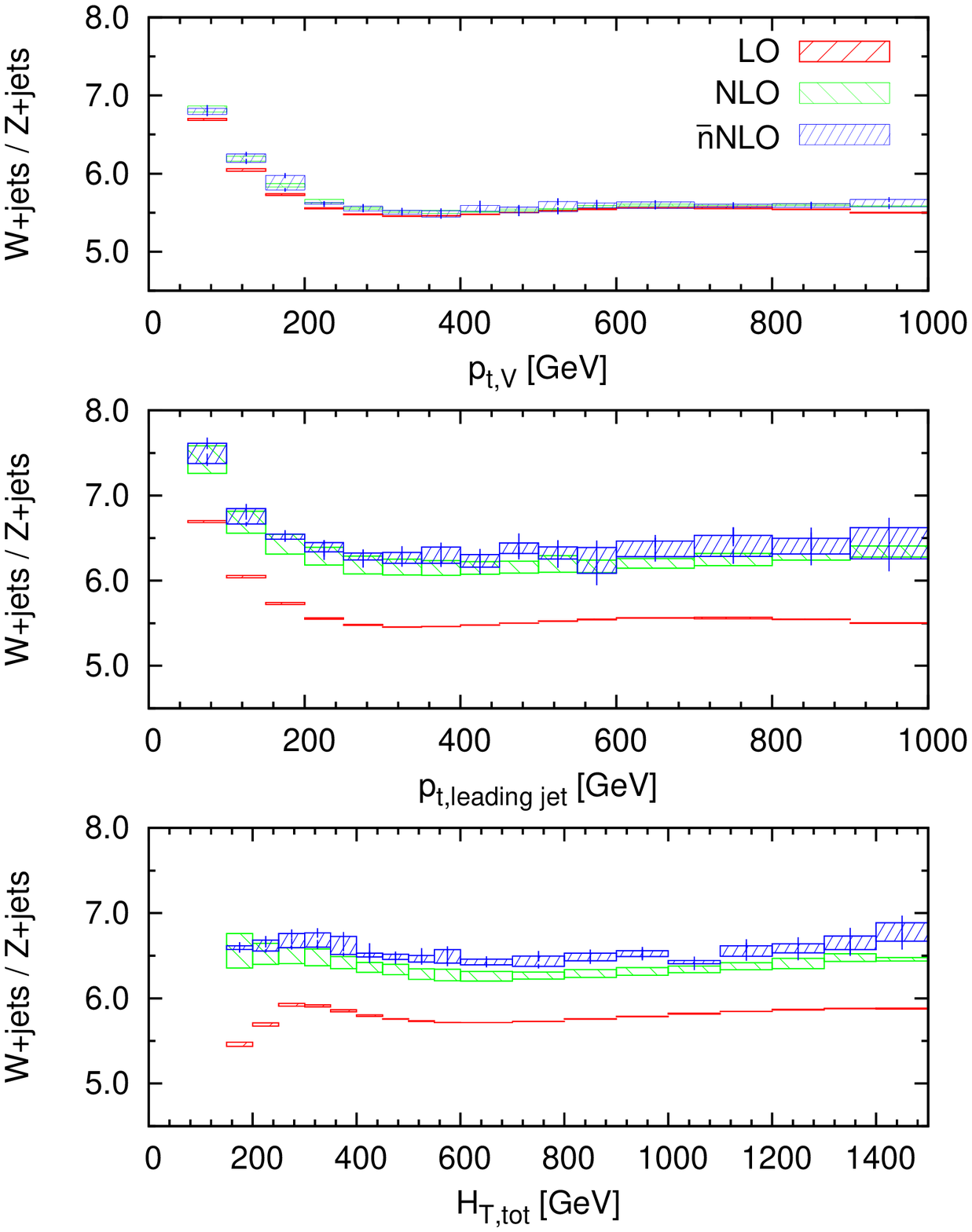}
  \includegraphics[width=0.5\columnwidth]{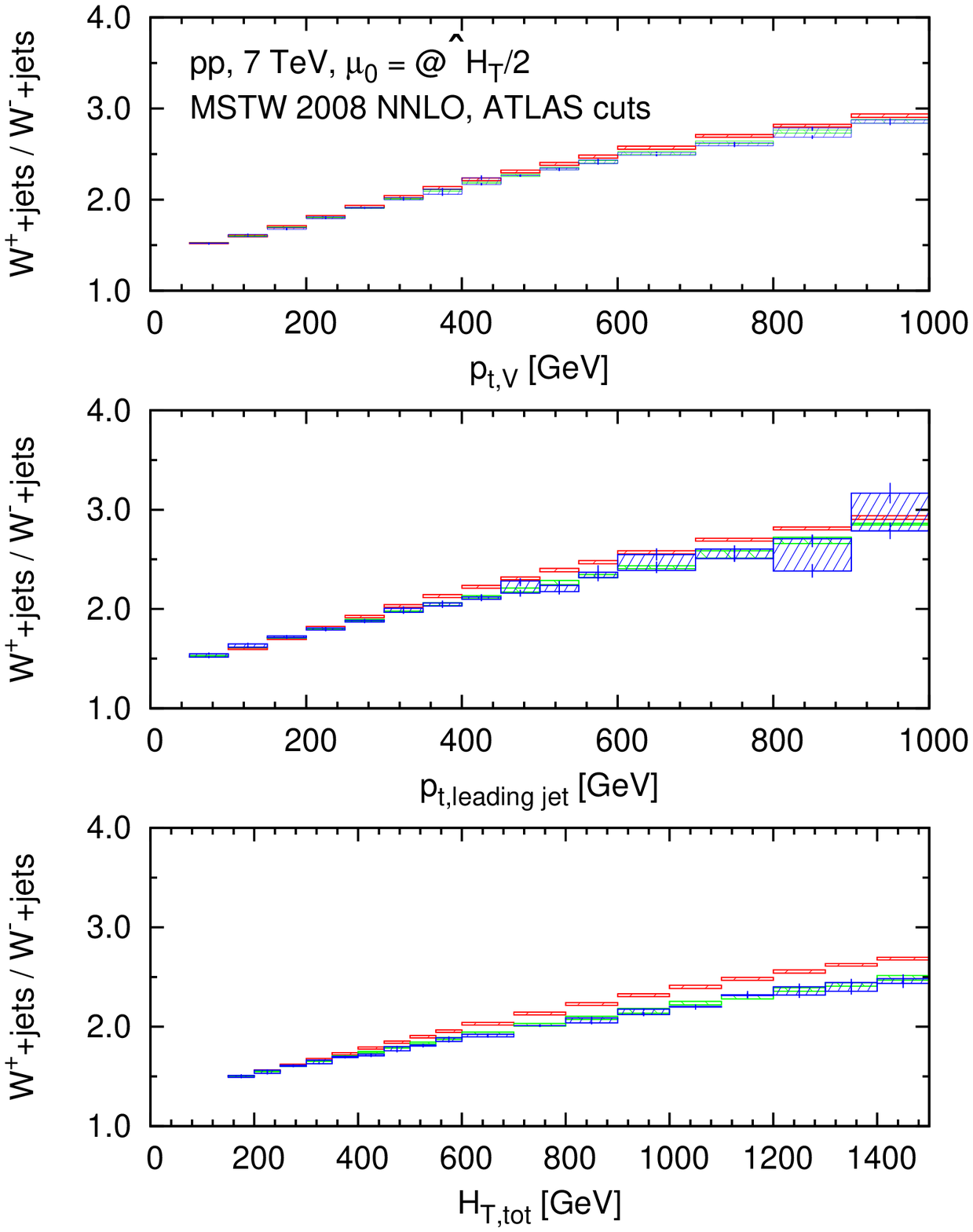}
  \caption{
  Ratios of the differential distributions of $p_{T,V}$, $p_{T,\text{leading
  jet}}$, $H_{T,\jet}$ and  $H_{T,\tot}$ at parton level at LO, NLO and \nNLO.
  The bands correspond to varying $\mu_F=\mu_R$ by factors 1/2 and 2 around the
  central value from Eq.~(\ref{eq:mufmur}) or changing $R_\LS$ to 0.5 and
  1.5.
  The vertical lines show statistical errors of the \nNLO result.
  }
  \label{fig:ratios}
\end{figure}

The productions of $\rm W^{\pm}$+jets and Z+jets bear obviously a number of
common features. It is therefore interesting to use the distributions discussed
in Sec.~\ref{sec:wjets} and \ref{sec:zjets}, to form ratios of W+jets/Z+jets and
W$^+$+jets/W$^-$+jets.\footnote{As mentioned earlier, whenever we write
``W+jets'' we really mean the sum: W$^+$+jets + W$^-$+jets.}
From the theory point of view, such ratios have a number of potential advantages
as the dependencies on kinematic variables, scales and PDFs, cancel to a large
extent. They are also motivated experimentally since they should be almost free
of uncertainties coming from luminosity measurement, jet energy scale or
unfolding.

In Fig.~\ref{fig:ratios} (left) we show the W+jets/Z+jets ratios as functions of
the $p_T$ of the electroweak boson, $p_T$ of the leading jet and $H_{T,\tot}$.
The bands correspond to varying the renormalization and factorization scale by
factors 1/2 and 2, simultaneously for the numerator and the denominator, hence
they indicate scale dependence of the ratios.
We see that the results are flat, regardless of the order, except for the
range below 200 GeV, where the difference between the masses of W and Z is still
non-negligible compared to their transverse momenta.
 
As shown in~\cite{Malik:2013kba}, the production of W+jets at high $p_T$ at LO
occurs predominantly via $gu$ and $gd$, for W$^+$ and W$^-$, respectively,
whereas for Z+jets it involves both $gu$ and $gd$ channels in
approximately equal proportions.  Therefore, the parton luminosities in the
numerator and in the denominator yield a very similar $x$ dependence, leading to
flat ratios.
Similar mechanism is at work for the subdominant $q\bar q$ channel where
W$^+$+jets is produced via $u\bar d$ and W$^-$+jets via $d\bar u$,
whereas Z+jets production involves the $u\bar u$ and $d\bar d$ 
channels.
The $f_{\bar u}$ and $f_{\bar d}$ parton distribution functions are, however,
very close to each other, which again results in a similar $x$-dependence in the
numerator and in the denominator, and hence the flat W+jets/Z+jets ratios.
The exact value of the LO ratios from Fig.~\ref{fig:ratios} (left), above 200
GeV, is an overall effect of different decay channels of the W and Z bosons and
different cuts used for the two processes.

The situation looks very similar at NLO, where the new production channels,
$qq$ and $gg$ are the same for the two processes and therefore do not
change the results qualitatively.
Quantitatively, as expected, since the new NLO channels only moderately modify
the $p_{T,V}$ distributions, the corresponding NLO ratio from
Fig.~\ref{fig:ratios} (left) receives no corrections at this order. On the
contrary, these new NLO channels and topologies contribute significantly to the
$p_{T,\text{ leading jet}}$ and $H_{T,\tot}$ distributions (cf.
Figs.~\ref{fig:wjets-allobs1} and \ref{fig:zjets-allobs1}), which, through
different cuts on the W and Z bosons decay products, lead to an increase of the
ratios.

Finally, the \nNLO results from Fig.~\ref{fig:ratios} (left) show that the
$\order{\alpha_\text{EW}\alpha_s^3}$ corrections have at most a few percent
effect on the ratios, to be compared with up to 50\% effect for separate
distributions from Figs.~\ref{fig:wjets-allobs1} and \ref{fig:zjets-allobs1}.
On the top of that, the renormalization and factorization scale dependence,
indicated by the widths of the bands shown in Fig.~\ref{fig:ratios}, is
very weak.

Fig.~\ref{fig:ratios} (right) shows the W$^+$+jets/W$^-$+jets ratios and the
observations made above largely apply also here. In particular, the NLO and
\nNLO corrections are small and the dependence on the renormalization and
factorization scale is significantly lower than for each distribution
separately. The W$^+$+jets/W$^-$+jets ratios, however, are not flat because they
involve the $f_u/f_d$ PDF ratio, which is an increasing function of~$x$ and
large $x$ values correspond to the tails of the distributions. 
This observation is consistent with findings from~\cite{Malik:2013kba}.

Altogether, the low sensitivity to higher order QCD corrections and weak
dependence on the factorization and renormalization scales make the ratios from
Fig.~\ref{fig:ratios} very good observables with great potential to be used in
precision studies.

\section{Conclusion}
\label{sec:conclusion}

We presented the study of W+jets and Z+jets production at the approximate
next-to-next-to leading order in QCD. We focused on the center of mass energy
$\sqrt{s} = 7 \TeV$ and we adopted the fiducial volumes of
~\cite{Aad:2012en,Aad:2013ysa}, which allowed for a direct comparison of our
results to the available ATLAS data.
We used the LoopSim method to merge NLO samples of different multiplicity
obtained from \mcfm~and from \bh+\sherpa~in order to compute the dominant
part of the NNLO corrections for jet observables at high $p_T$.
Our predictions, referred to as \nNLO, are expected to be accurate in the
regions of phase space dominated by new channels and new topologies appearing at
NLO. This corresponds to the large $p_T$/$H_T$ regions of distributions.

We found that, for both processes, the leading jet $p_T$ distribution comes
under control at \nNLO, with the scale uncertainty being reduced by up to 70\%
and the result staying within the NLO band. On the contrary, the $H_T$-type
observables still receive significant corrections at \nNLO, of the order of 50\%
with respect to NLO. 
We also checked that, as expected, for the distributions of the $p_T$ of the
electroweak bosons, leptons and missing $E_T$, which do not exhibit a large
K-factor at NLO, the \nNLO result does not bring any significant correction.

For both W+jets and Z+jets processes, we compared our \nNLO results to the
experimental data from ATLAS. For $p_{T, \text{leading jet}}$, the agreement
between data and theory is comparable to that at NLO. As mentioned above, the
\nNLO result exhibits, however, much smaller theoretical uncertainty.
Nevertheless, the statistical errors of the data are still too
large at high $p_T$ to favour one prediction of over the other.
On the other side, the $H_{T,\tot}$ and $H_{T,\jet}$ distributions at \nNLO
agree much better with the data than NLO. As discussed in Sec.~\ref{sec:wjets},
this is due to the fact that the \nNLO result includes V+2j configurations at
NLO and V+3j configurations at LO, which represent a sizable contribution at
high $H_T$. 
It would be interesting to investigate whether the situation improves
further at \nnNLO, where the V+4jets configurations would be included. This is
left for future work.

The code used in our study is publicly available at: 
\begin{center}
\url{https://loopsim.hepforge.org}.
\end{center}
It contains the LoopSim library together with the interfaces to \mcfm~{\sc 6.6}
and to \roo~ntuples. The \bh+\sherpa~ntuples will be made available soon and the next
version of \mcfm~{\sc 6.7} will include several extra features that will make it
easier to interface with LoopSim.

\section*{\normalsize Acknowledgement}

We thank Gavin Salam for numerous discussions during this work and for
subsequent comments on the manuscript.
We are grateful to Ulla Blumenschein and Joey Huston for useful
conversations, clarifying a number of experimental issues, and for
critical reading of the manuscript.
We acknowledge valuable discussions with Stefano Camarda and Nicolas Meric at
various stages of this work.
We are grateful to Graeme Watt for pointing us to relevant findings concerning
ratios of W and Z cross sections.
We thank Alberto Guffanti and Pavel Storovoitov for smooth collaboration on the
extensions of \mcfm, and the \mcfm~authors for including these new features in
the next release.
We thank the \bh+\sherpa~authors for providing us with the \roo~ntuples.
We acknowledge correspondence with Alexander Paramonov concerning experimental
details of the W+jets results from ATLAS.
DM's work was supported by the Research Executive Agency~(REA) of the European
Union under the Grant Agreement number PITN-GA-2010-264564 (LHCPhenoNet).


\end{document}